\documentclass[10pt, oneside]{article}
\pdfoutput=1
\usepackage{hyperref}
\usepackage{enumitem} 
\usepackage{easylist}  	
\usepackage{graphicx}			
\usepackage{amssymb}
\usepackage[usenames]{color}
\usepackage{epsf}
\usepackage[usenames,dvipsnames]{pstricks}
\usepackage{epsfig}
\usepackage{pst-grad} 
\usepackage{url}
\usepackage{wrapfig}
\usepackage{float} 
\newenvironment{s-enumerate}{
\begin{enumerate}
  \setlength{\itemsep}{1pt}
  \setlength{\parskip}{0pt}
  \setlength{\parsep}{0pt}
}{\end{enumerate}}

\let\OLDthebibliography\thebibliography
\renewcommand\thebibliography[1]{
  \OLDthebibliography{#1}
  \setlength{\parskip}{.5ex}
  \setlength{\itemsep}{0pt plus 0.3ex}
}
\title{The Ameyalli-Rule: Logical Universality in a 2D Cellular Automaton}
\author{Jos\'e Manuel G\'omez Soto%
\thanks{jmgomez@uaz.edu.mx, http://matematicas.reduaz.mx/$\sim$jmgomez}%
\hspace{2ex}{\it \small Universidad Aut\'onoma de Zacatecas.}\\
\hspace{2ex}{\it \small  Unidad Acad\'emica de Matem\'aticas. Zacatecas, Zac. M\'exico.}\\
Andrew Wuensche%
\thanks{andy@ddlab.org,  http://www.ddlab.org
}
\hspace{2ex}{\it \small Discrete Dynamics Lab.}\\
\\
{\normalsize Dedicated to the memory of John Horton Conway}\\ 
{\normalsize 1937-2020}
}

\begin{document}

\maketitle

\vspace{-3ex}
\begin{abstract}

\noindent We present a new spontaneously emergent glider-gun in a 2D
Cellular Automaton and build the logical gates NOT, AND and OR
required for logical universality.  The Ameyalli-rule is not based on
survival/birth logic but depends on 102 isotropic neighborhood groups
making an iso-rule, which can drive an interactive input-frequency
histogram for visualising iso-group activity and dependent functions
for filtering and mutation.  Neutral inputs relative to
logical gates are identified which provide an idealized
striped-down form of the iso-rule.

\end{abstract}

\begin{center}
{\it keywords: cellular automata, iso-rule, glider-gun, logical gates, universality}
\end{center}

\section{Introduction}
\label{Introduction}

The Ameyalli-rule\footnote{Ameyalli means a spring of running water in
Nahuatl, a language spoken in central Mexico.} is a new result that
continues our search for 2D Cellular Automata (CA) with glider-guns
and eaters capable of logical universality.  Since the publication of
Conway’s Game-of-Life\cite{Gardner1970} with its survival/birth s23/b3
logic, many other ``Life-Like'' survival/birth combinations have been
examined\cite{Eppstein2010} but none seem to have come close\cite{Minondo2021}
to achieving the complexity of behaviour of the Game-of-Life itself.

To study the basic principles of CA universal computation in a more
general context, glider-guns and logical gates have been demonstrated
outside survival/birth ``Life-Like'' constraints, but still within
isotropic rule-space where all possible flips/spins of a neghborhood
pattern give the same input. Isotropic rules are preferable to enact
logical universality because their dynamics have no directional bias
and computational machinery operates in any orientation.
Examples include the 3-value 7-neighbour hexagonal
Spiral-rule\cite{Adamatzky&Wuensche2006}, and for a binary 2d Moore
neigborhood, the Sapin-rule\cite{Sapin2004}, and four rules in our own published
results demonstrating logical universality ---
the first was the anisotropic X-Rule\cite{Gomez2015}, followed by the
isotropic Precursor-rule\cite{Gomez2017}, the
Sayab-rule\cite{Gomez2018} and the Variant-rule\cite{Gomez2020}.

\begin{figure}[htb] 
  \begin{center}
\begin{minipage}[c]{.7\linewidth}
(1) \includegraphics[width=.12\linewidth,bb=92 76 149 132, clip=]{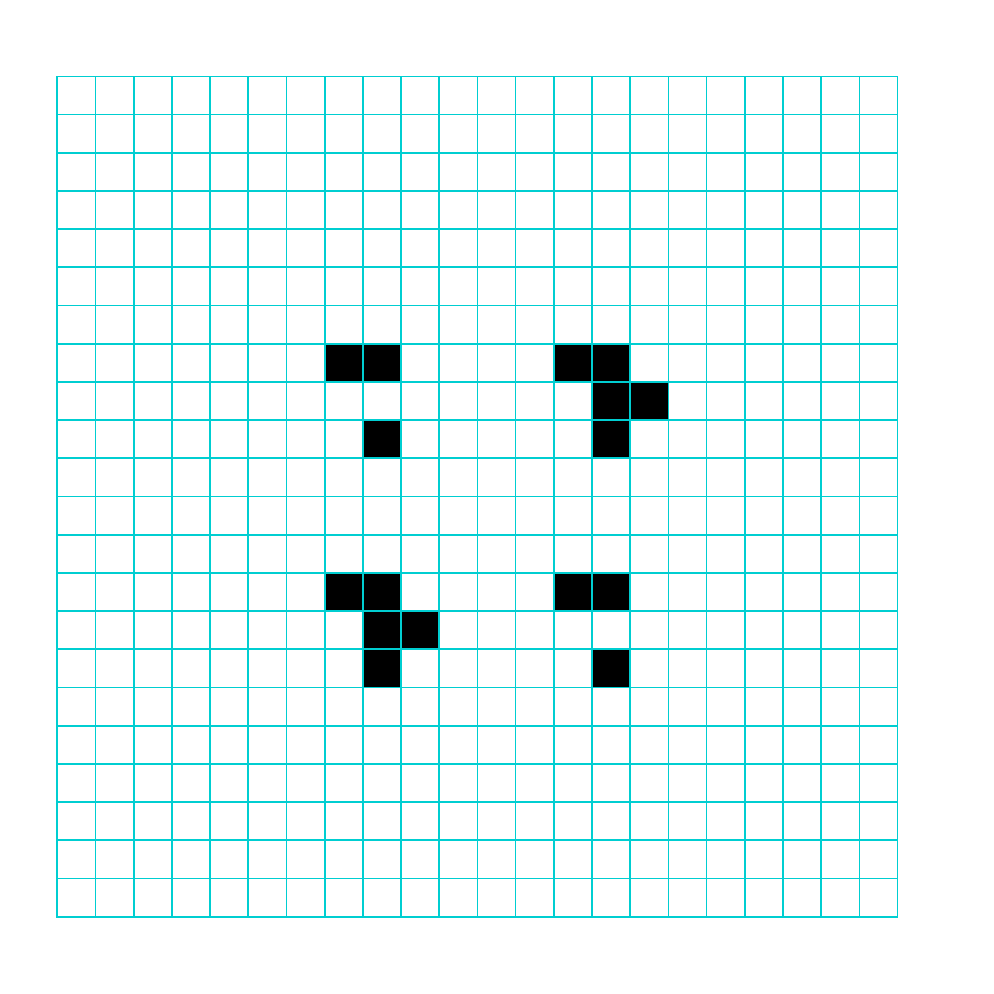}
\hfill
(2) \includegraphics[width=.12\linewidth,bb=158 76 215 132, clip=]{am_pdf-figs/am-gldr}
\hfill
(1) \includegraphics[width=.12\linewidth,bb=82 76 139 132, clip=]{am_pdf-figs/am-gldr}
\hfill
(2) \includegraphics[width=.12\linewidth,bb=71 142 128 198, clip=]{am_pdf-figs/am-gldr}
\end{minipage}
\end{center}
\vspace{-4ex}
\caption[Ameyalli-rule glider]
        {\textsf{
            The Ameyalli-rule 2-phase glider moving East at a speed of c/2.\\
            4 time-steps are shown across a fixed background with phases labelled (1) and (2).            
\label{am-gldr}
        }}
        \vspace{-2ex}
\end{figure}

\begin{figure}[htb] 
\begin{minipage}[c]{1\linewidth}
\includegraphics[width=1\linewidth,bb=323 76 1050 796, clip=]{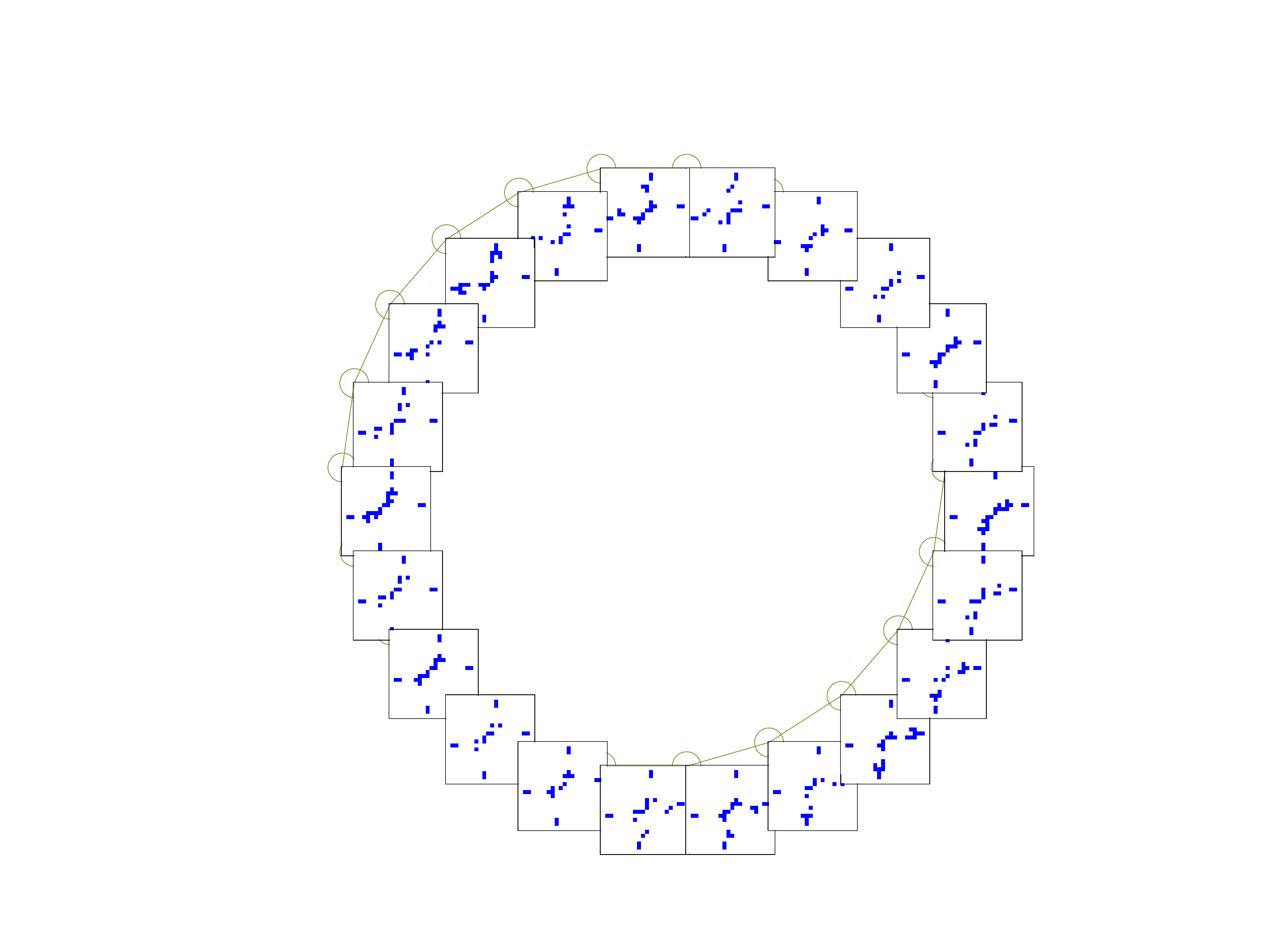}
\end{minipage}
\begin{minipage}[c]{1\linewidth}
  \vspace{-75ex}\hspace{35ex}\includegraphics[width=.34\linewidth,bb=0 -1 196 192, clip=]{am_pdf-figs/am-gg22} 
\end{minipage}
\vspace{-5ex}
\caption[The glider-gun attractor cycle]
{\textsf{
The Ameyalli period 22 glider-gun shown as an attractor cycle\cite{Wuensche92,EDD}
with time running clockwise.
Here the glider-gun is confined to its central core by type-B eaters. Inset is  a detail of
the initial time-step. 
\label{attractor-cycle}
}}
\vspace{-2ex}
\end{figure}

These rules were found from a short-list\cite{Gomez2015,Gomez2017}
within an input-entropy scatter-plot\cite{Wuensche99,EDD} sample of
93000+ isotropic rules, which classify rule-space between order,
chaos and complexity.  The input-entropy criteria in this sample
follow ``Life-Like'' constraints to the extent that the rules are
binary, isotropic, with a 3$\times$3 Moore neighborhood ---
\raisebox{-.8ex}
{\includegraphics[height=3ex, bb=10 14 37 40,clip=]{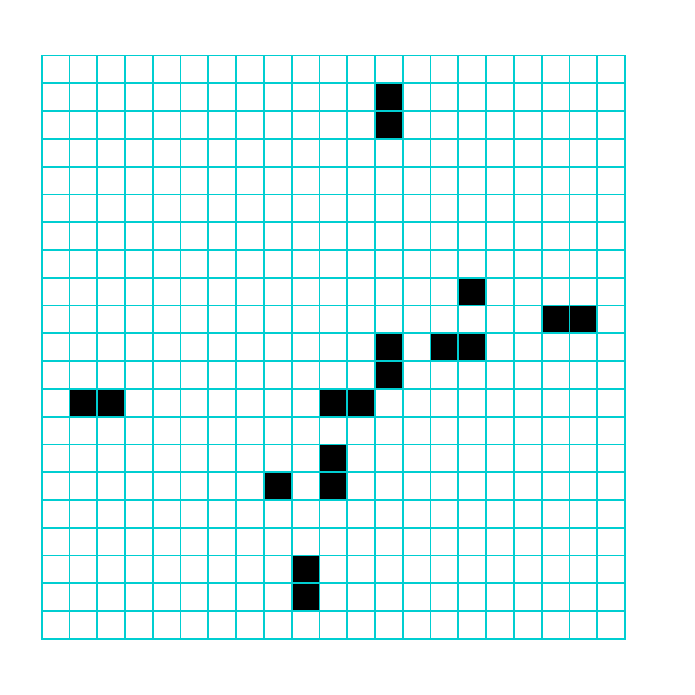}}
--- and with the $\lambda$~parameter, the
density of 1s in the full 512 rule-table, similar to the
Game-of-Life where $\lambda=0.273$.  The short-list consists of rules
that feature emergent gliders within the ordered zone of the plot.
The Ameyalli-rule was found from the same
short-list. Its 2-phase orthogonal emergent glider --- one type found to date
--- is shown in figure~\ref{am-gldr}.
Its glider-gun (figures~\ref{attractor-cycle} and \ref{glidergun}) also emerges
spontaneously from a random initial state in a similar way to the
Sapin-rule\cite{Sapin2004}, the Sayab-rule\cite{Gomez2018}, and the
Spiral-rule\cite{Adamatzky&Wuensche2006}, but with a lower
probability, whereas the glider-guns for the other rules listed above,
including the Game-of-Life, require careful construction.

\begin{figure}[htb] 
\begin{minipage}[c]{1\linewidth}
\includegraphics[width=1\linewidth,bb=8 13 824 25, clip=]{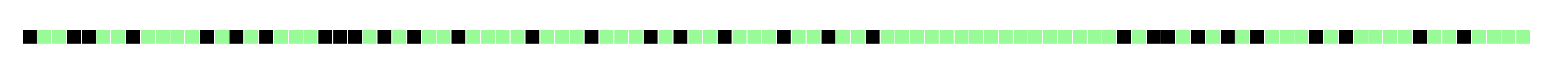}\\[-2ex] 
\includegraphics[width=1\linewidth,bb=8 13 824 25, clip=]{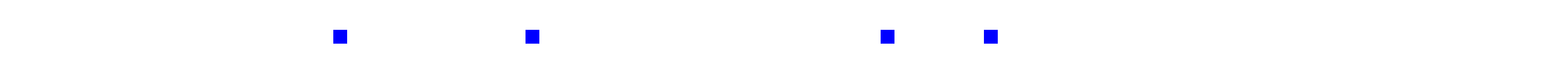}\\[-.5ex] 
\textsf{\small \phantom{xxxxxxxxxxxxxxxxxxx}eater-B\phantom{xxxxxxxxxxxxxxxxxxxx}eater-A}
\end{minipage}
\vspace{-3ex}
\caption[Ameyalli-rule ]
        {\textsf{
The Ameyalli 102-bit iso-rule shown as a DDLab graphic,
indexed from 101-0 (left-right) 
--- the positions of mutants from the
rule originally found are indicated below the graphic in blue. 
For eater-A, indices 43 and 36 were flipped from 1 to 0. 
For eater-B, indices 78 and 67 were flipped from 0 to 1. 
The iso-rule expressed in hexadecimal is
26 42 a3 a9 08 8a 44 90 00 0b 54 50 90.
\label{am-iso}
}}
\end{figure}

The Ameyalli-rule is most efficiently defined and mutated
as a 102-bit iso-rule\cite{Wuensche2021},
where any mutation conserves isotropy.  An in-depth definition
of isotropic rules, iso-rules based on iso-groups, including
alternative notations --- the full rule-table, symmetry
classes, iso-rule prototypes, and the iso-rule in hexadecimal --- are
presented in section~\ref{The Ameyalli-Rule definition}. These
notations relate to DDLab\cite{Wuensche-DDLab} and can be redefined
for Golly\cite{Golly}.  Figure~\ref{am-iso} shows the Ameyalli
iso-rule, and the mutations that were made to create the eaters A and B.

It turn out that a large proportion, 60/102, of Ameyalli iso-rule
inputs are neutral with respect its glider-gun/eater system
(section~\ref{The idealized iso-rule}) so this system
can be equivalently generated by a vast family of mutant
iso-rules. The identities of the Ameyalli and other logically universal
rules are centered on their glider-gun/eater systems, which is
best represented by a stripped-down ``idealized'' version of the
iso-rule with all neutral inputs set to zero.

\enlargethispage{3ex}
The paper is organised into the following sections:
(\ref{The glider-gun, gliders, eaters, and collisions}) describes
the glider-gun, gliders, eaters, and collisions,
(\ref{Logical Universality}) outlines logical universality,
\ref{Logical Gates} demonstrates the logical gates,
(\ref{The Ameyalli-Rule definition}) defines the iso-rule,
(\ref{iso-rule input-frequency-histogram (IFH)}) gives methods for iso-rule activity
by the input-frequency histogram (IFH), for filtering and mutation,
and for idealising the iso-rule, and
\ref{Summary and Discussion} is a summary and discussion of the issues.
\clearpage

\section{The glider-gun, gliders, eaters, and collisions}
\label{The glider-gun, gliders, eaters, and collisions}

\begin{figure}[htb] 
  \includegraphics[height=.63\linewidth,bb=77 126 511 532, clip=]{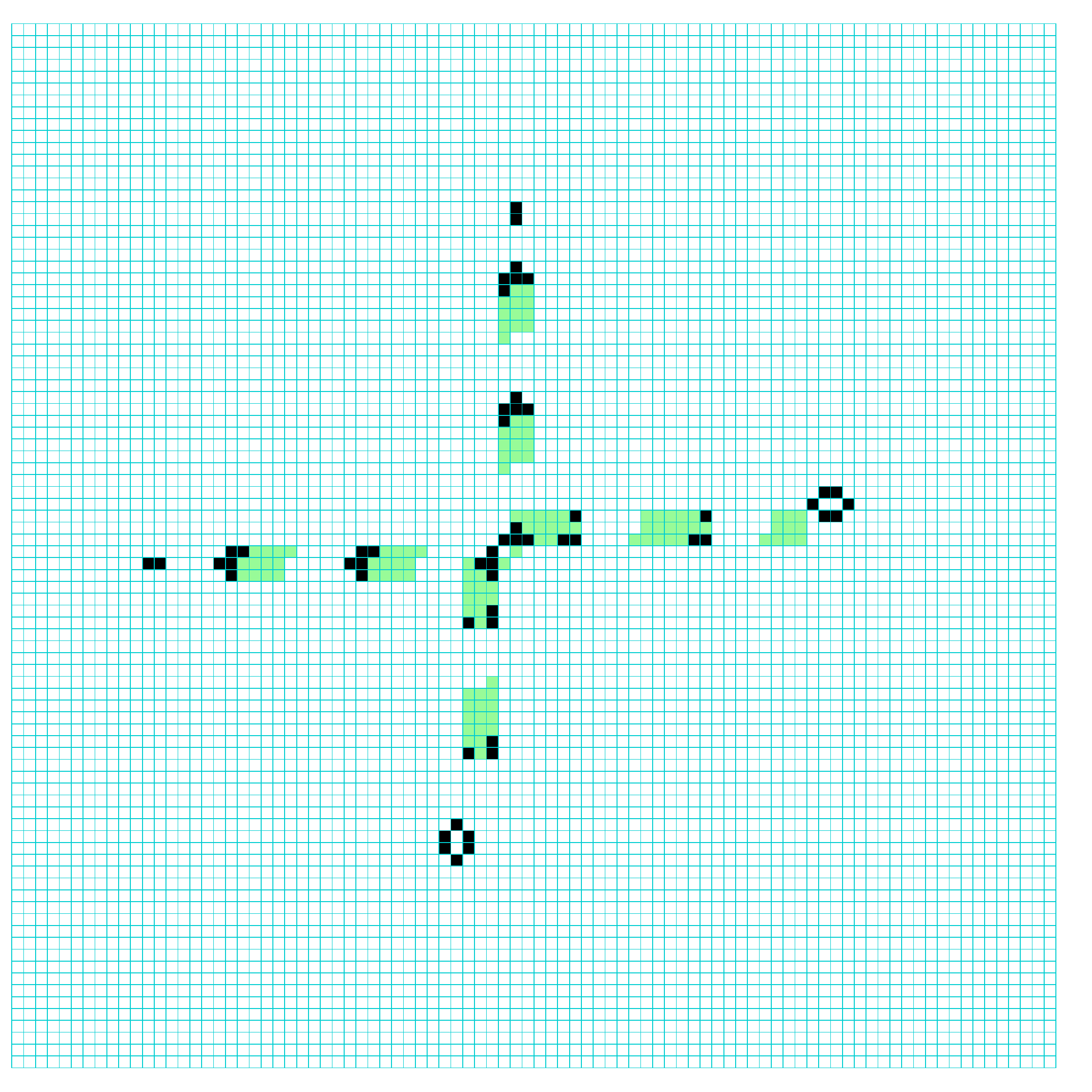} 
  \hfill
  \includegraphics[height=.63\linewidth]{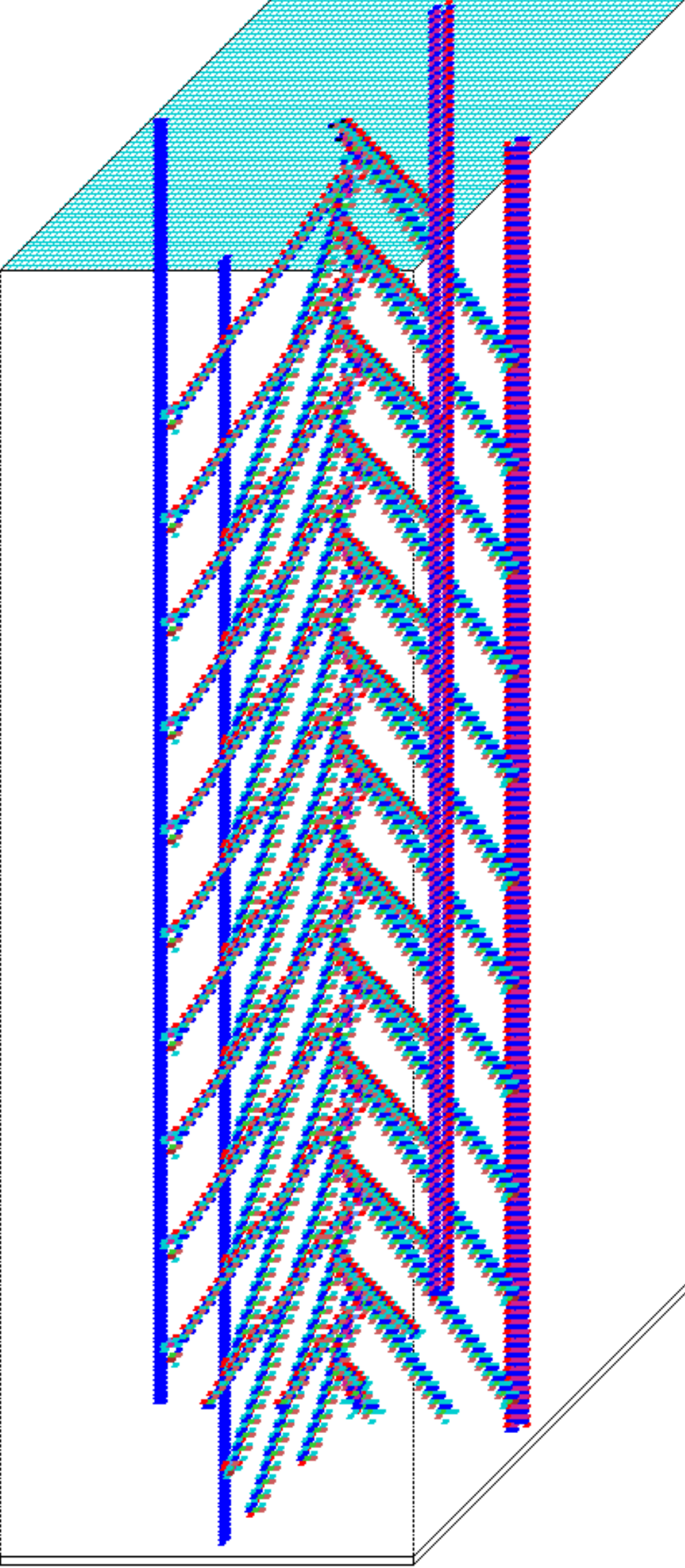} 
\vspace{-2ex}
\caption[The glider-gun in Ameyalli-Rule]
{\textsf{
    The Ameyalli glider-gun. $Left$: a snapshot with green trails indicating motion.
    $Right$: 274 time-step isometric with IFH colors (section~\ref{iso-rule input-frequency-histogram (IFH)}).
    2-phase orthogonal
    gliders are shot in 4 directions North/South/East/West with period 22 and speed of $c$/2, 
    where $c$ is the speed of light. The gliders spacing in the glider-stream is 11 cells.
    In this example North and West glider streams are stopped by
    the stationary eater B, East and South by the 2-period eater A. 
\label{glidergun}
}}
\end{figure}

The essential ingredients for a recipe to create CA logical universality
are gliders, glider-guns, eaters, and appropriate collisions.  

A glider is a special kind of oscillator, a mobile pattern that
recovers its form but in a displaced position, thus moving at a given
velocity. A rule with the ability to support a glider, together with a
stable eater, and a diversity of interactions between gliders and
eaters, provides the first hint of potential universality.

Although many rules can be found with these properties, the really
essential and most elusive ingredient is a glider-gun, a dynamic
structure that ejects gliders periodically into space. A glider-gun
can also be seen as an oscillator that adds to its form periodically
to shed gliders. If can be regarded as a
periodic attractor\cite{Wuensche99}, especially if confined by eaters
as in figure~\ref{attractor-cycle}.

In the case of the Ameyalli-rule, as already noted, a
glider-gun may emerge spontaneously from a random pattern --- only one type
has been observed so far. In other cases 
a glider-gun can sometimes be constructed from smaller components, as for
the very first glider-gun created by Gosper for the Game-of-Life. 
These glider-guns are such elaborate structures that the probability 
of their spontaneous emergence is negligible.

\begin{figure}[htb] 
\begin{minipage}[b]{.623\linewidth}
\includegraphics[width=.148\linewidth,bb=279 448 330 533, clip=]{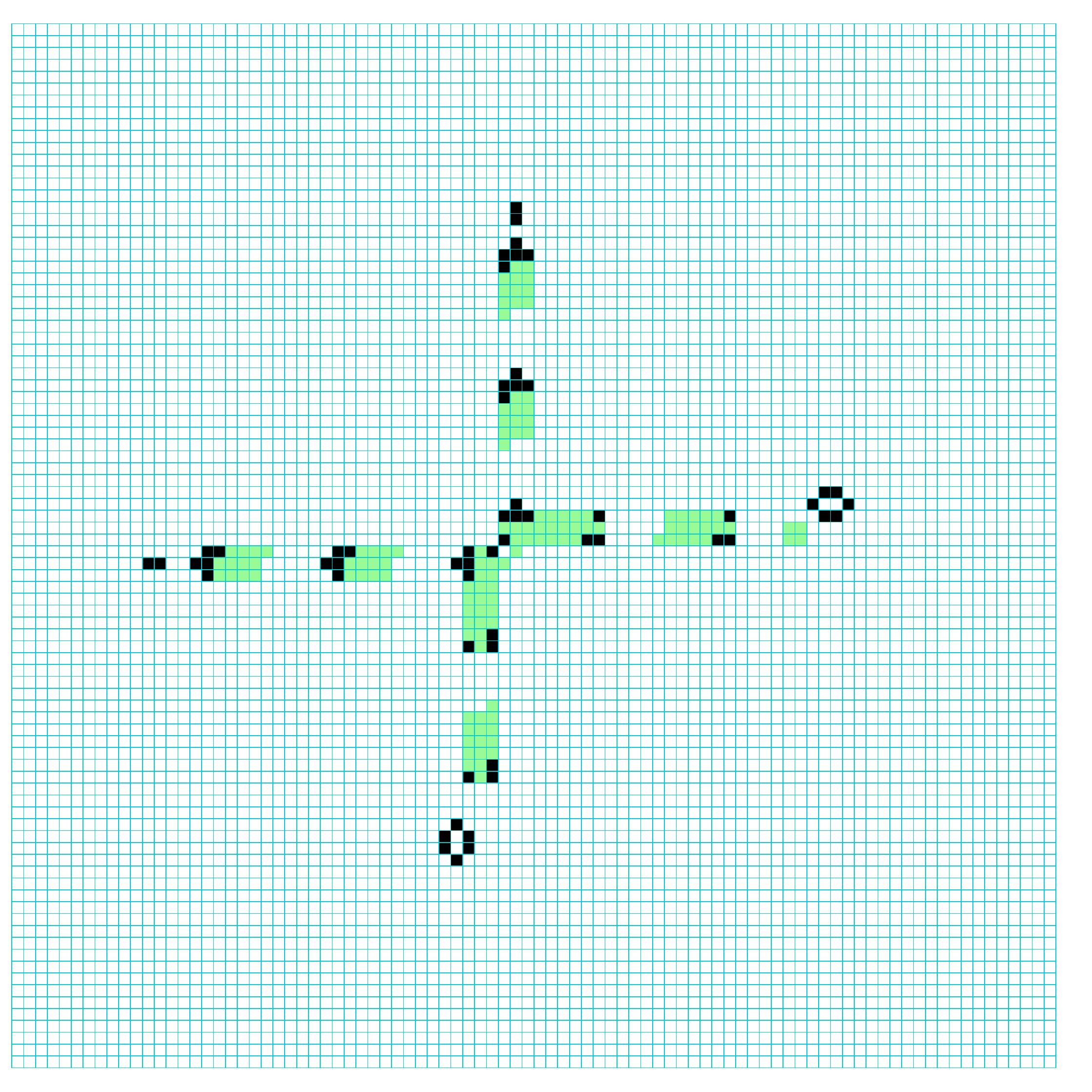}  \hfil 
\includegraphics[width=.148\linewidth,bb=279 448 330 533, clip=]{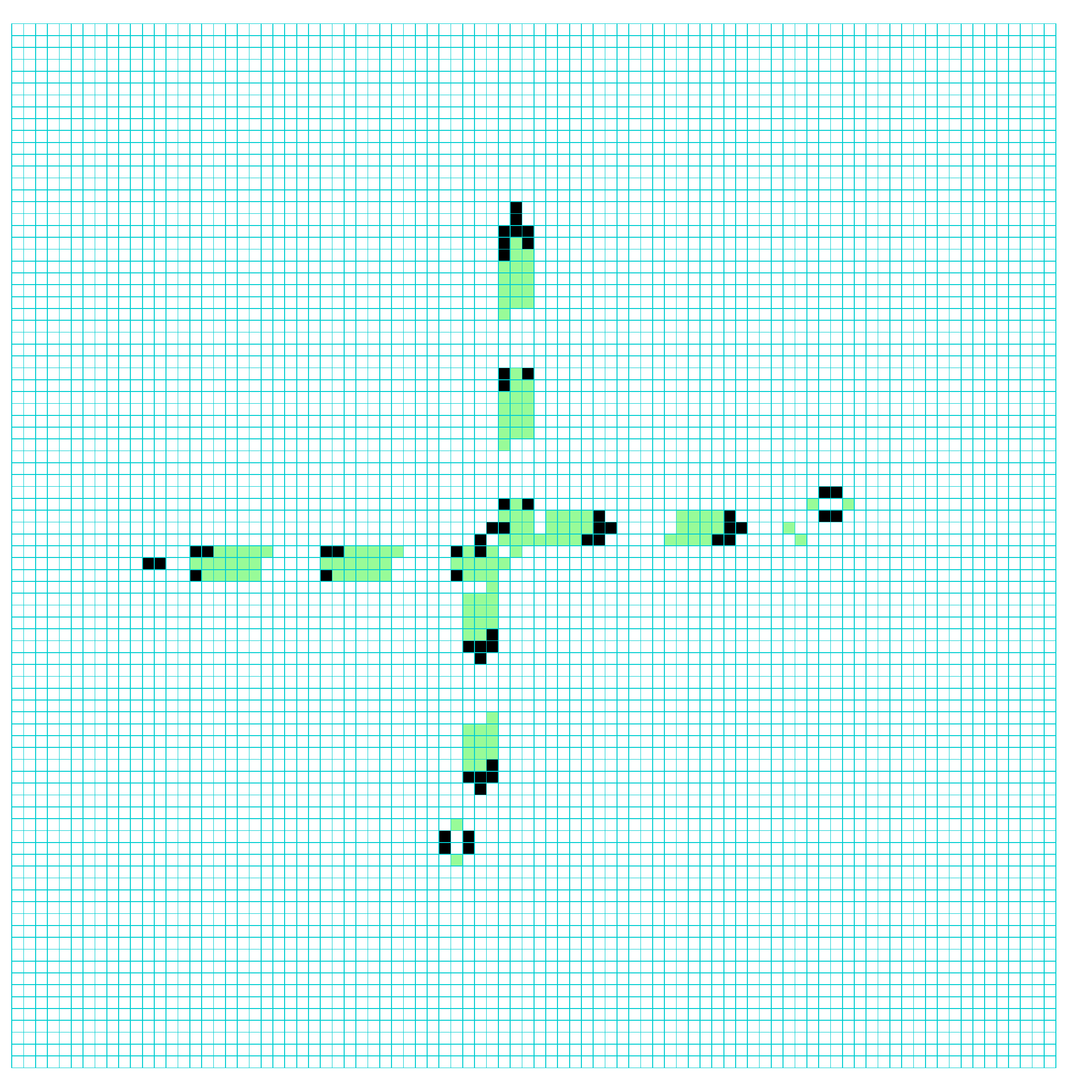}  \hfil 
\includegraphics[width=.148\linewidth,bb=279 448 330 533, clip=]{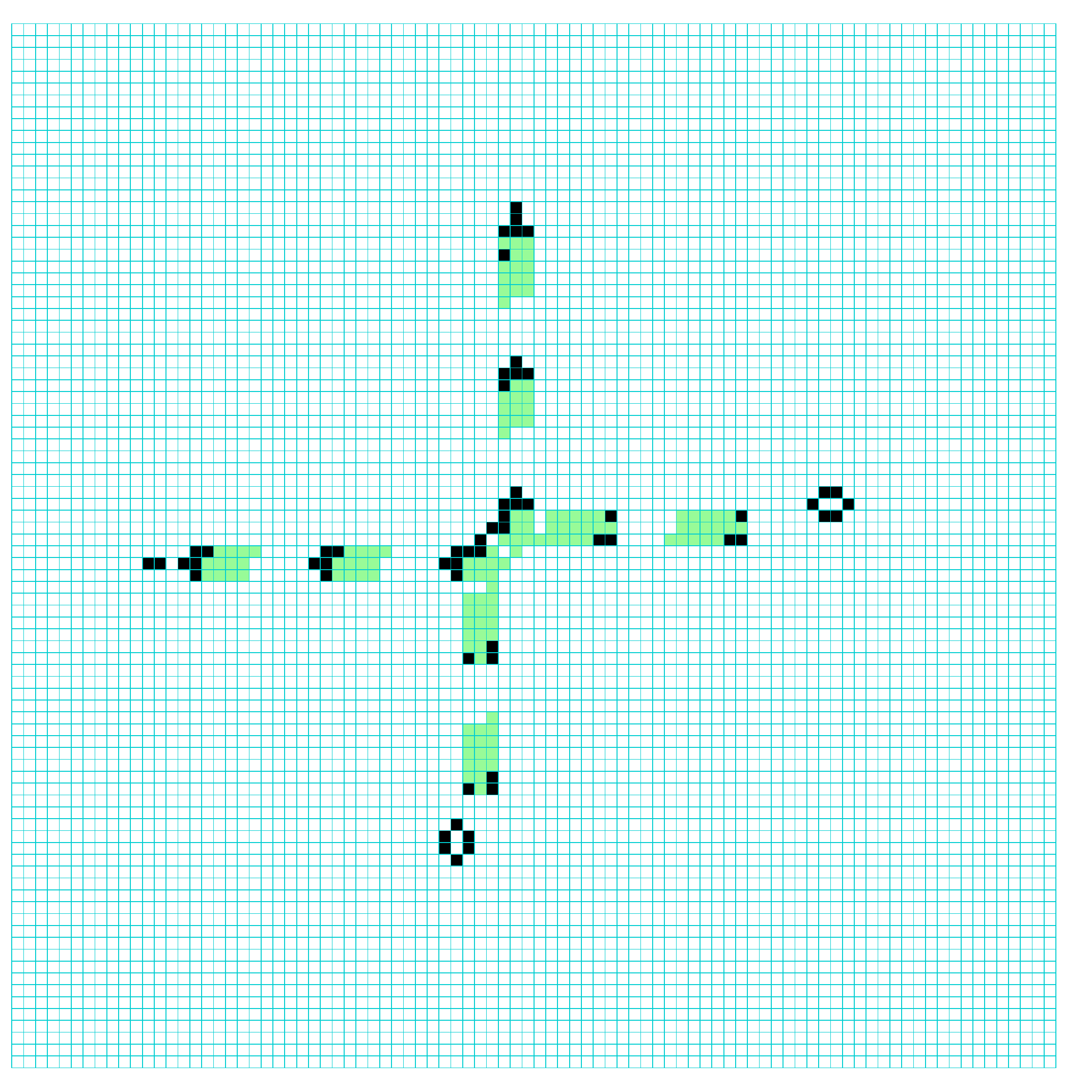}  \hfil 
\includegraphics[width=.148\linewidth,bb=279 448 330 533, clip=]{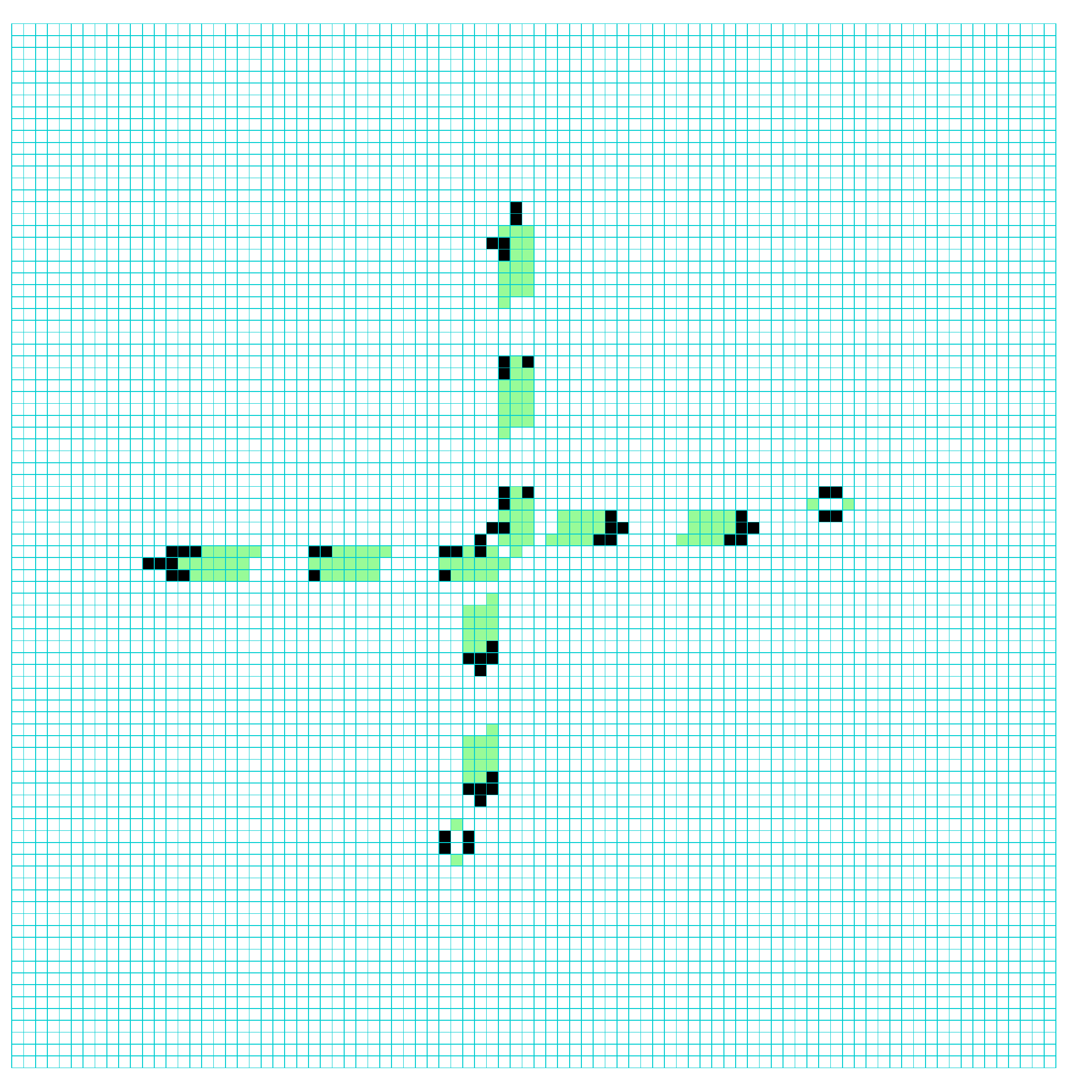}  \hfil 
\includegraphics[width=.148\linewidth,bb=279 448 330 533, clip=]{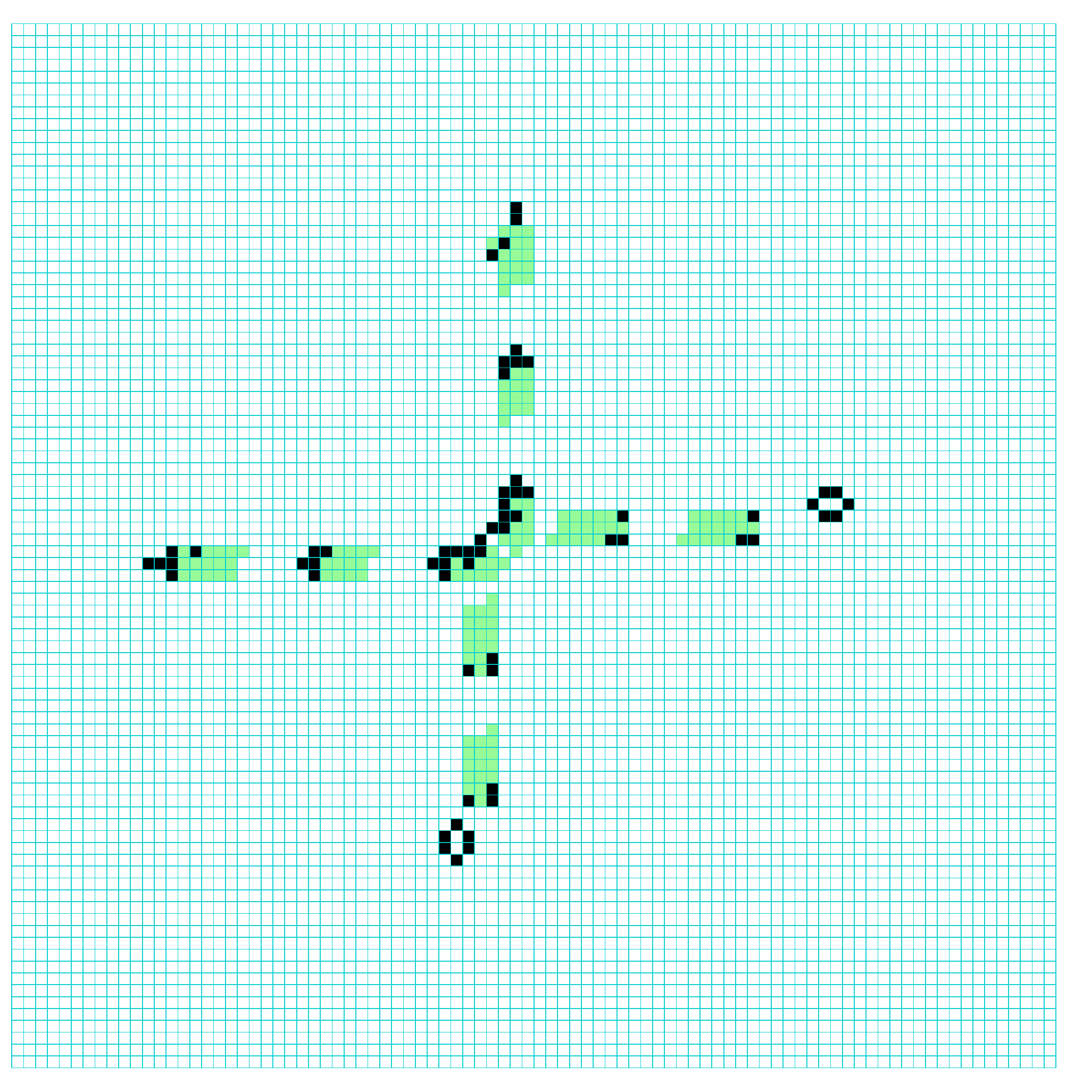}  \hfil 
\includegraphics[width=.148\linewidth,bb=279 448 330 533, clip=]{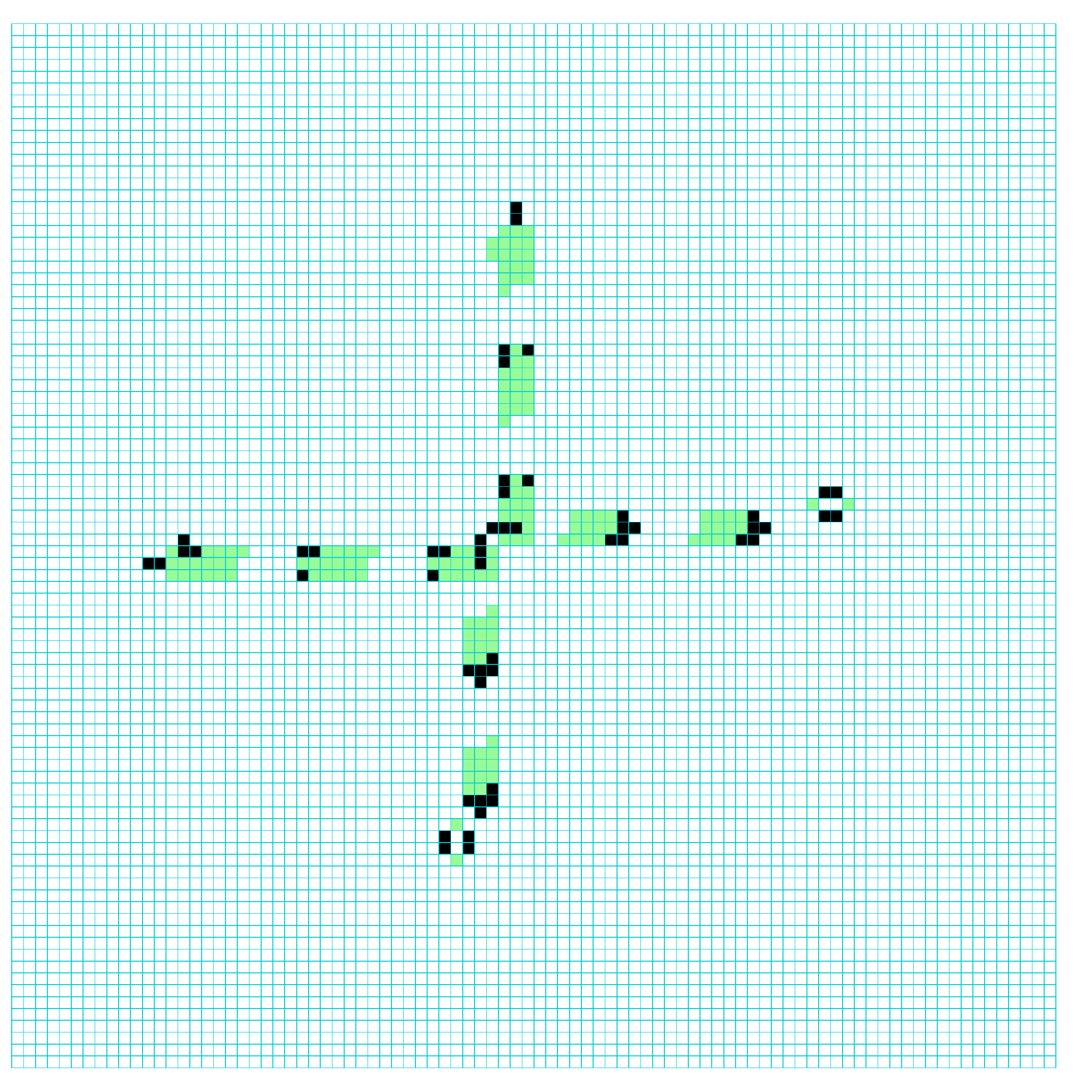} 
\end{minipage}
\phantom{x}
\begin{minipage}[b]{.30\linewidth} 
{\textsf{\small  
collision between a\\  
stationary eater-B and\\
a glider noving North}}\\
\end{minipage}\\[1ex]
\begin{minipage}[b]{.72\linewidth}
\includegraphics[width=.126\linewidth,bb=252 126 301 224, clip=]{am_pdf-figs/am-c5} \hfill 
\includegraphics[width=.126\linewidth,bb=252 126 301 224, clip=]{am_pdf-figs/am-c6} \hfill 
\includegraphics[width=.126\linewidth,bb=252 126 301 224, clip=]{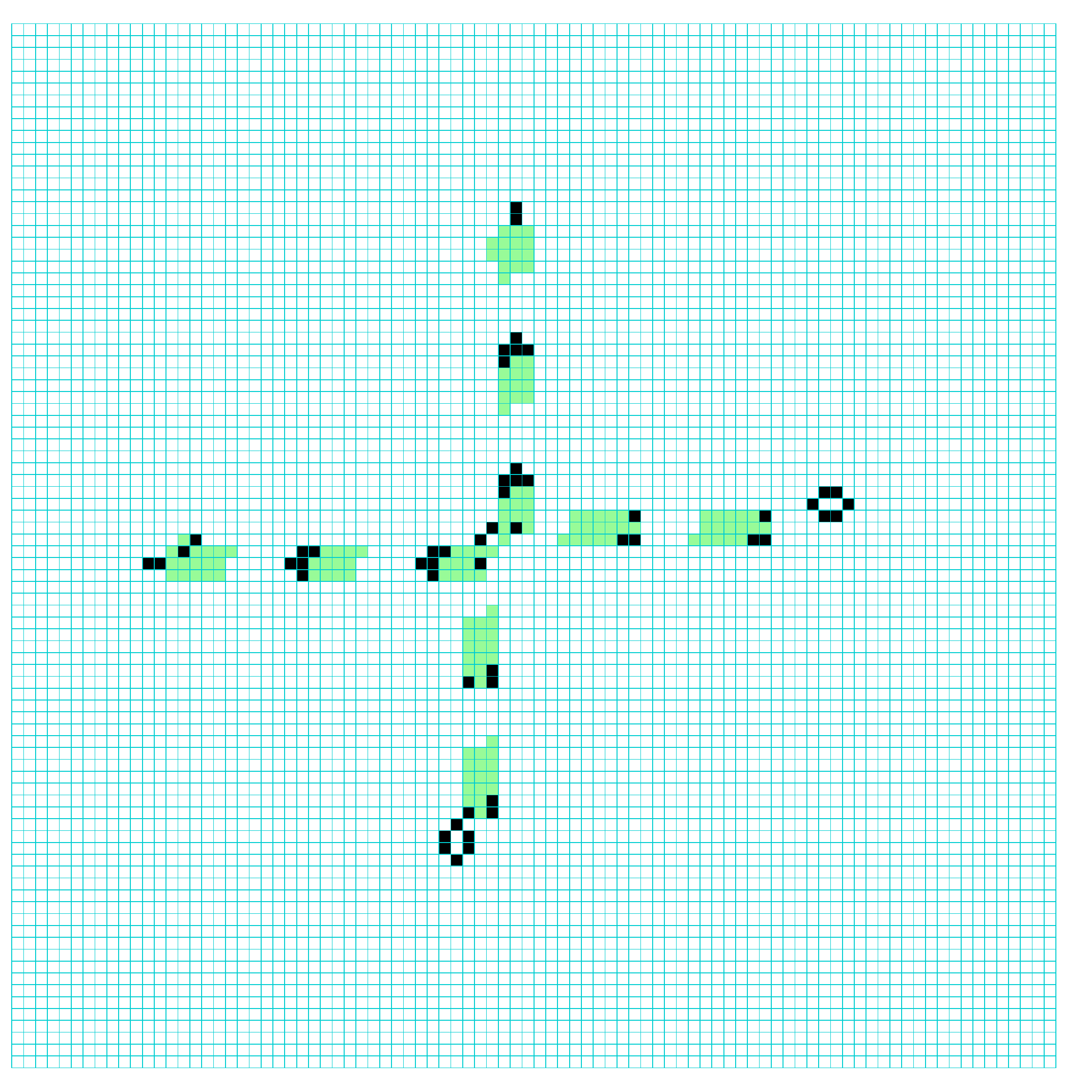} \hfill 
\includegraphics[width=.126\linewidth,bb=252 126 301 224, clip=]{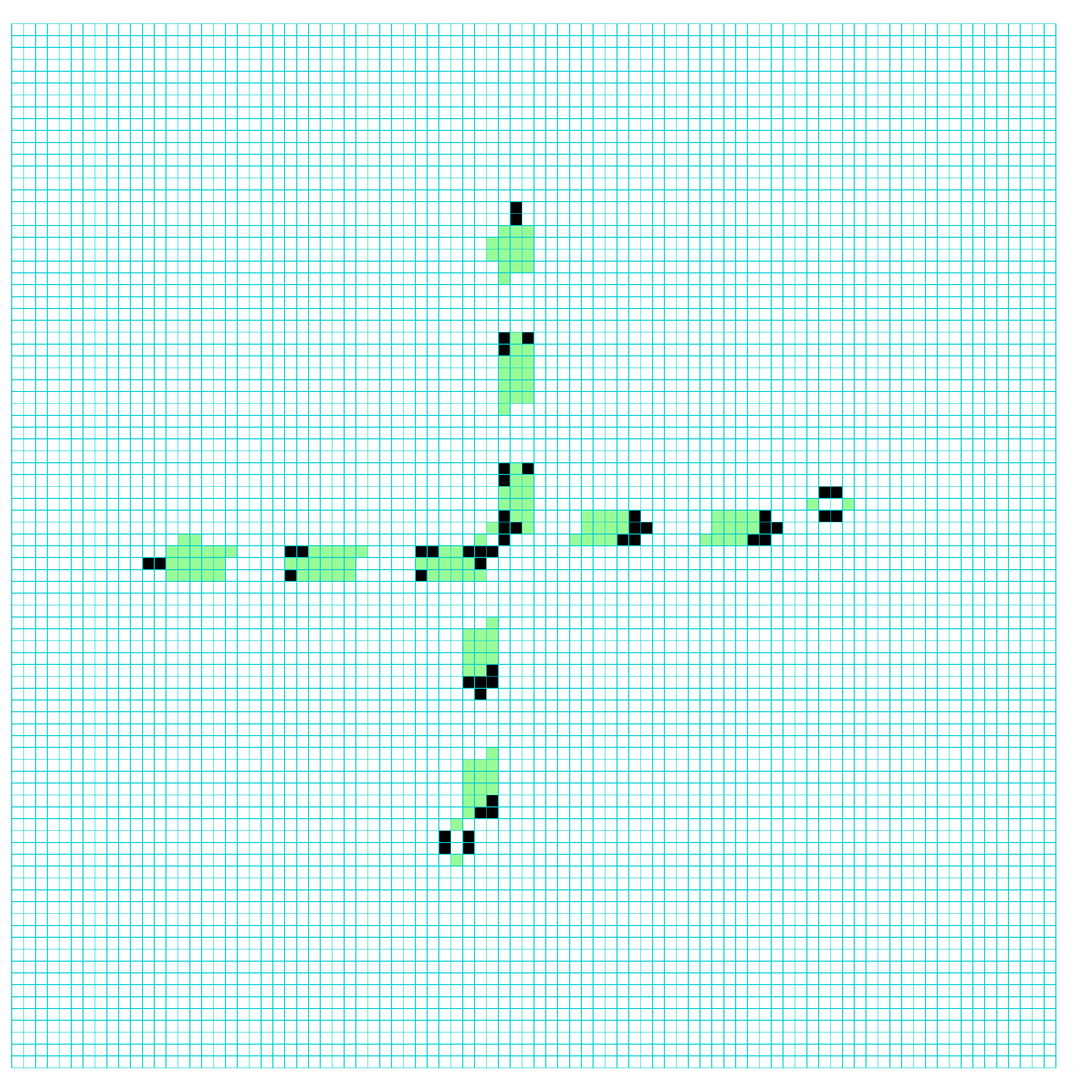} \hfill 
\includegraphics[width=.126\linewidth,bb=252 126 301 224, clip=]{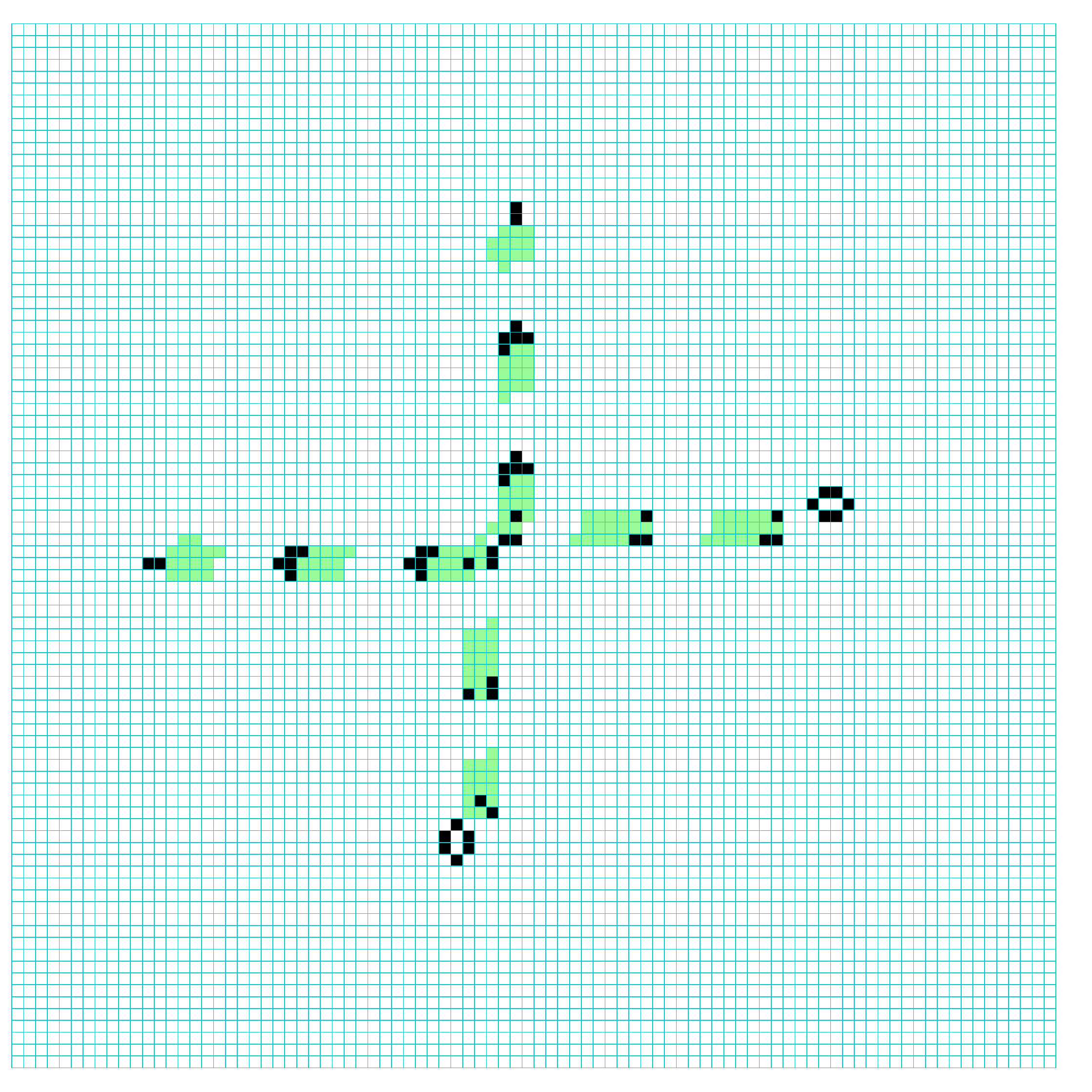} \hfill 
\includegraphics[width=.126\linewidth,bb=252 126 301 224, clip=]{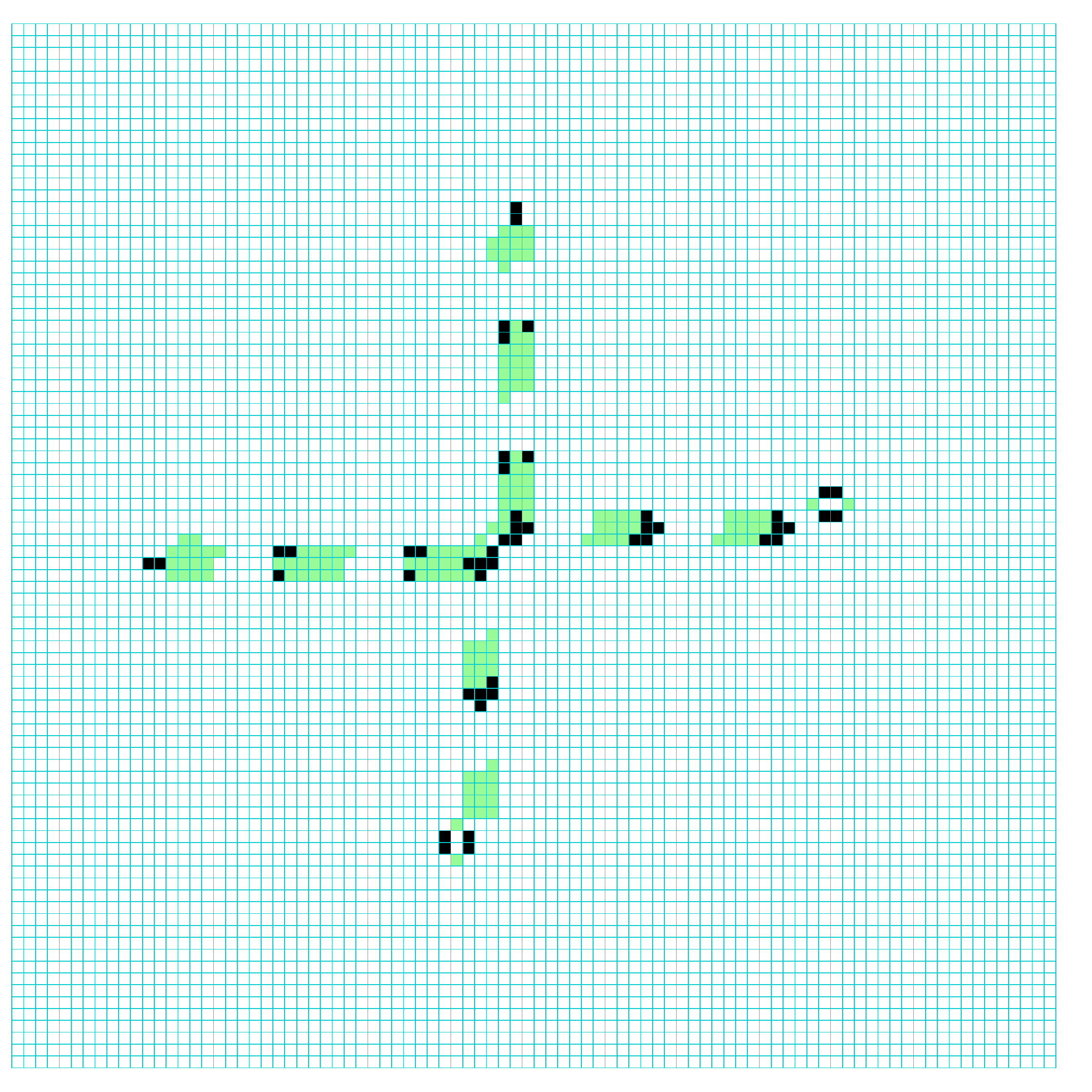} \hfill 
\includegraphics[width=.126\linewidth,bb=252 126 301 224, clip=]{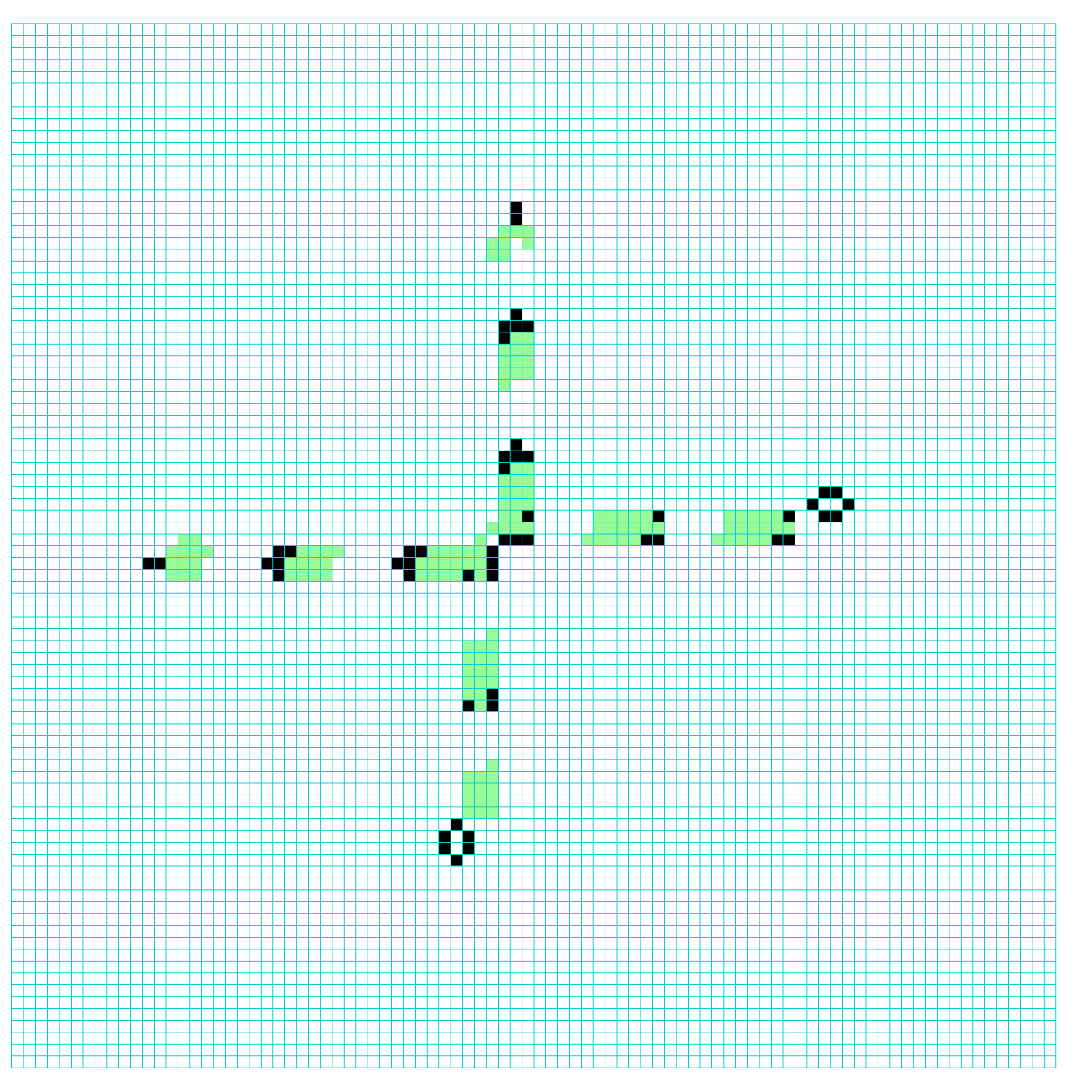} 
\end{minipage}
\phantom{x}
\begin{minipage}[b]{.25\linewidth}
{\textsf{\small  
collision between a\\  
2-phase eater-A and\\
a glider moving South}}\\
\end{minipage}
\vspace{-3ex}
\caption[2 types of eater]
{\textsf{
There are two types of eater, type-A oscillates with period 2, type-B is stationary. 
We illustrate above how a glider self-destructs when colliding with an eater.
If the collision dynamics are exactly coordinated in time and space, the eater
will re-establish its original configuration, and will be able to destroy the next
glider in a glider-gun-stream. Green trails indicate motion.
\label{eaters}
}}
\end{figure}

\begin{figure}[htb] 
\begin{center}
\begin{minipage}[b]{1\linewidth}
\includegraphics[width=.13\linewidth,bb=322 342 399 448, clip=]{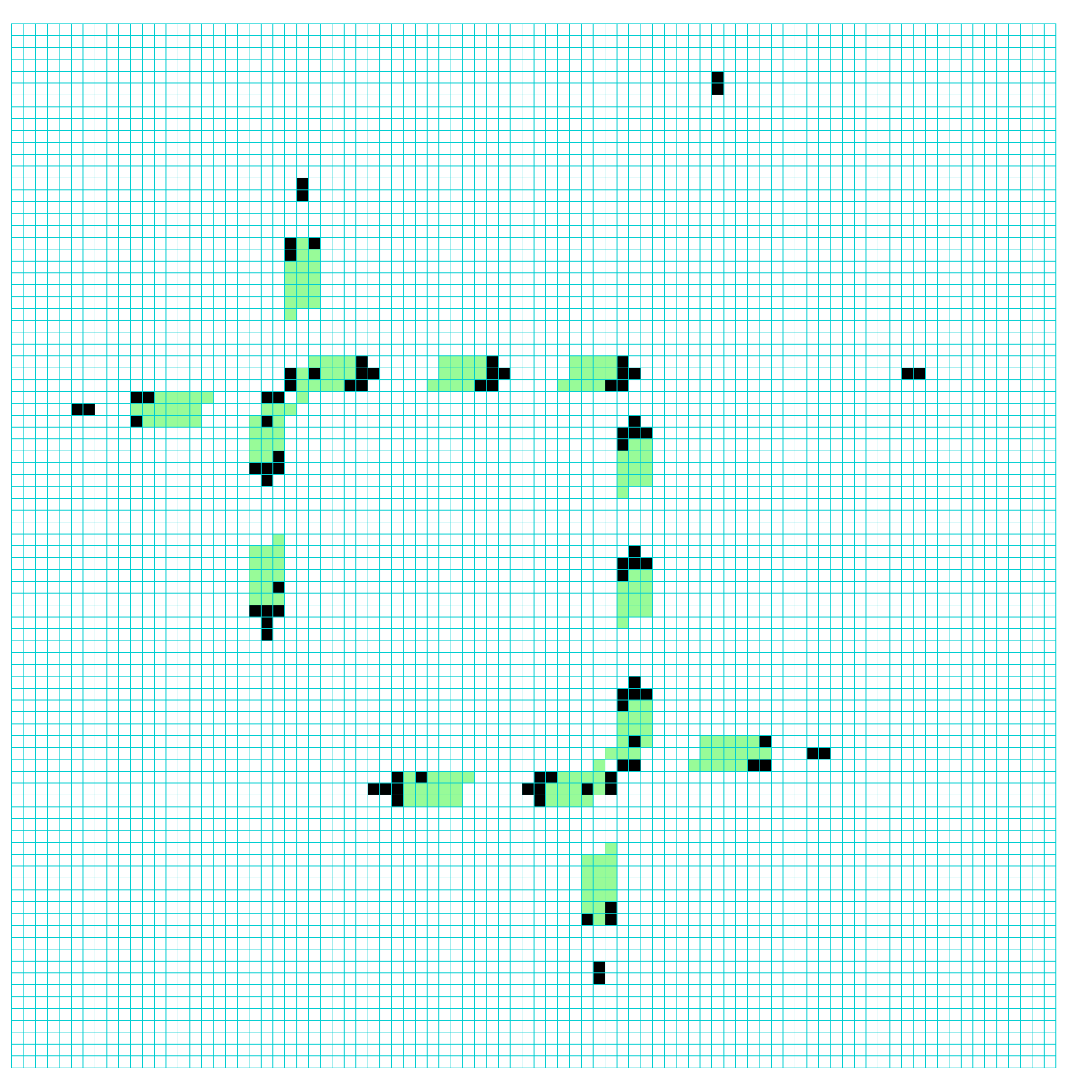} \hfill 
\includegraphics[width=.13\linewidth,bb=322 342 399 448, clip=]{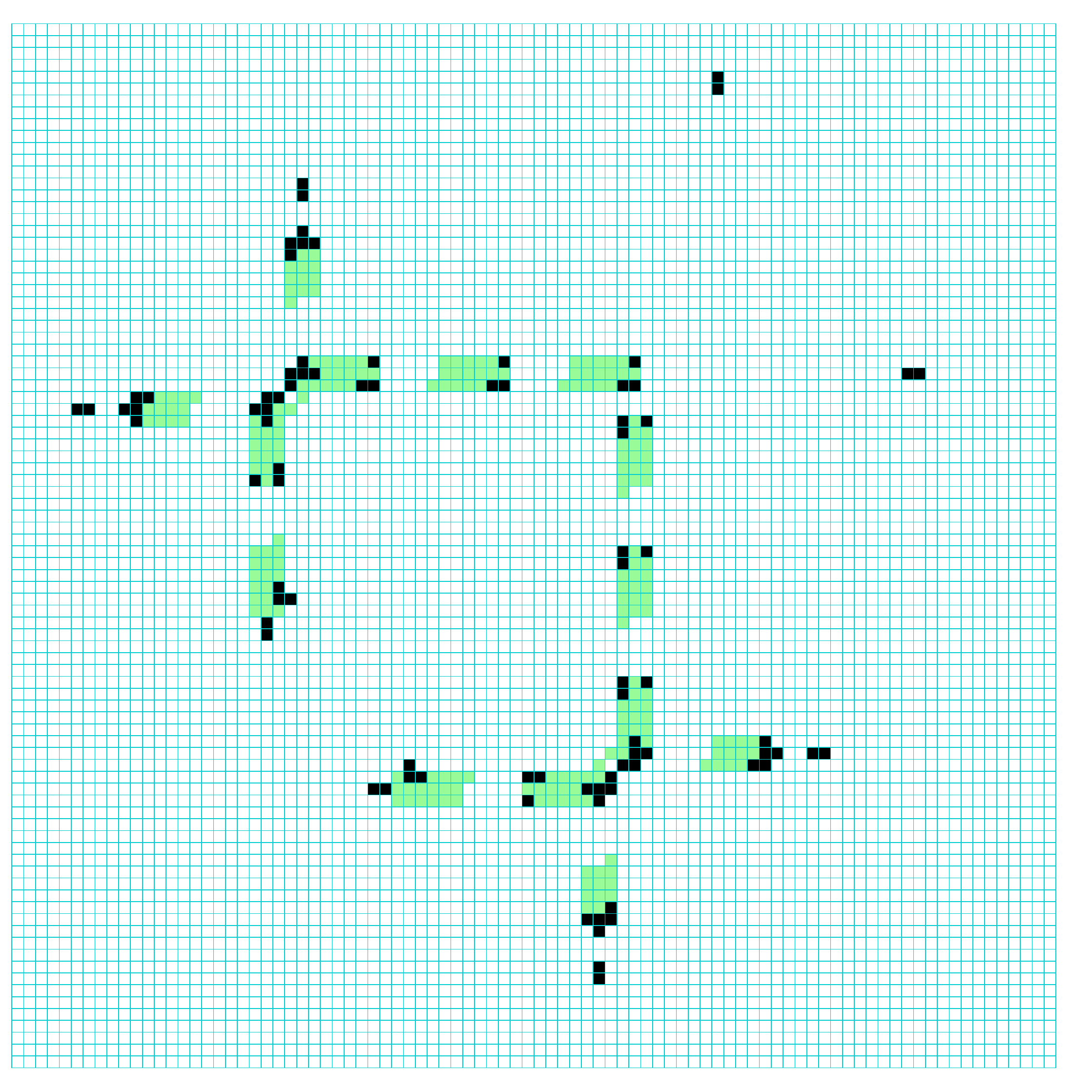} \hfill 
\includegraphics[width=.13\linewidth,bb=322 342 399 448, clip=]{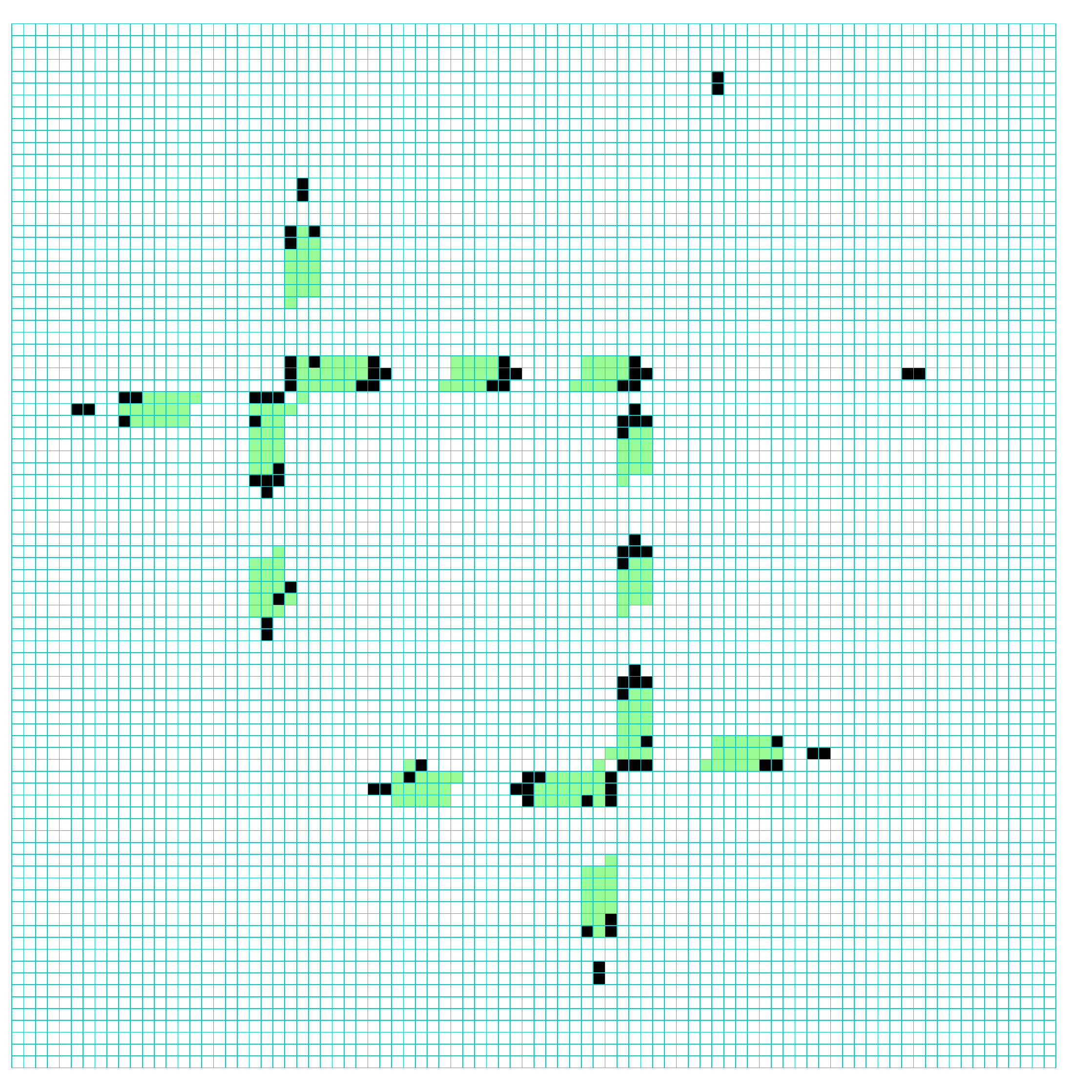} \hfill 
\includegraphics[width=.13\linewidth,bb=322 342 399 448, clip=]{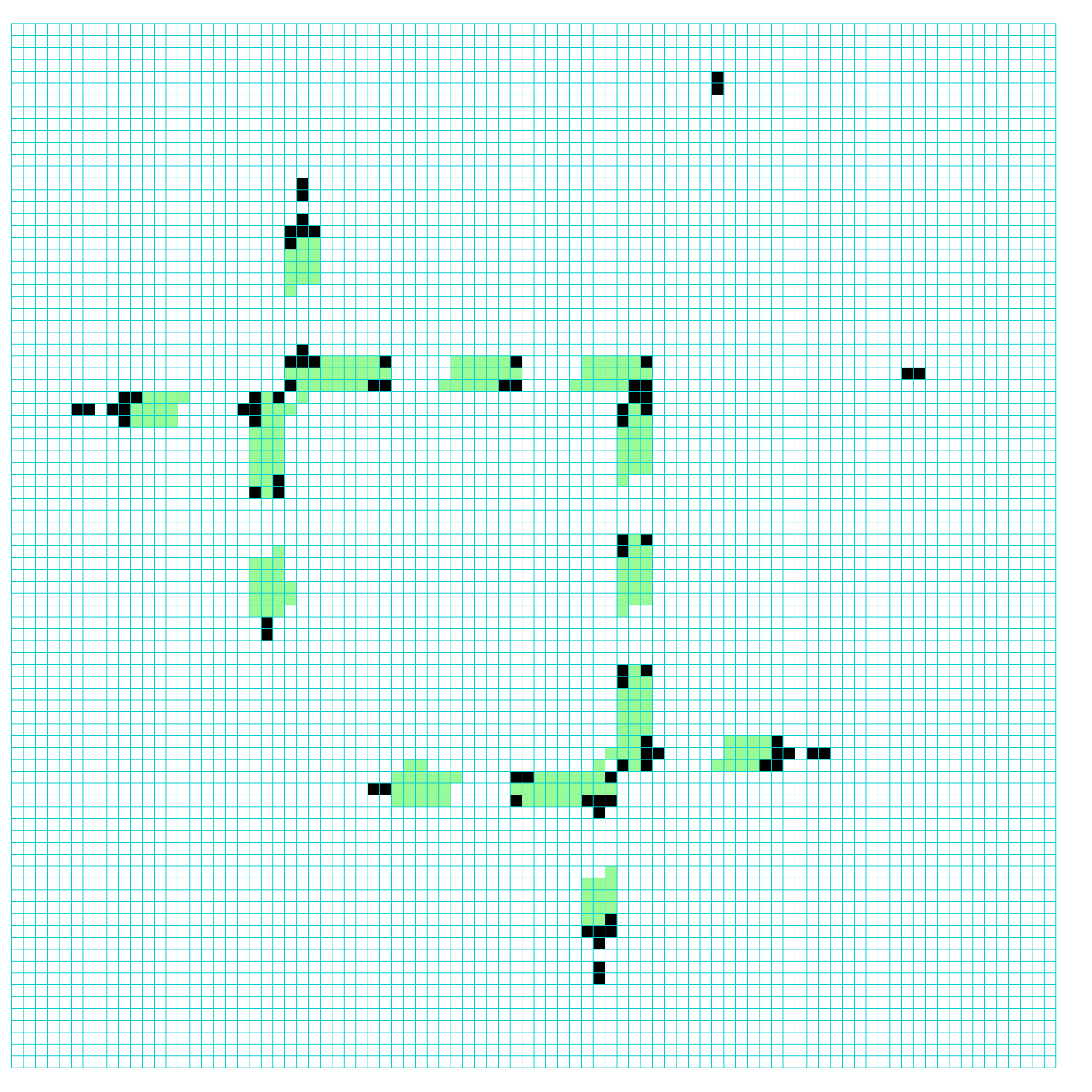} \hfill 
\includegraphics[width=.13\linewidth,bb=322 342 399 448, clip=]{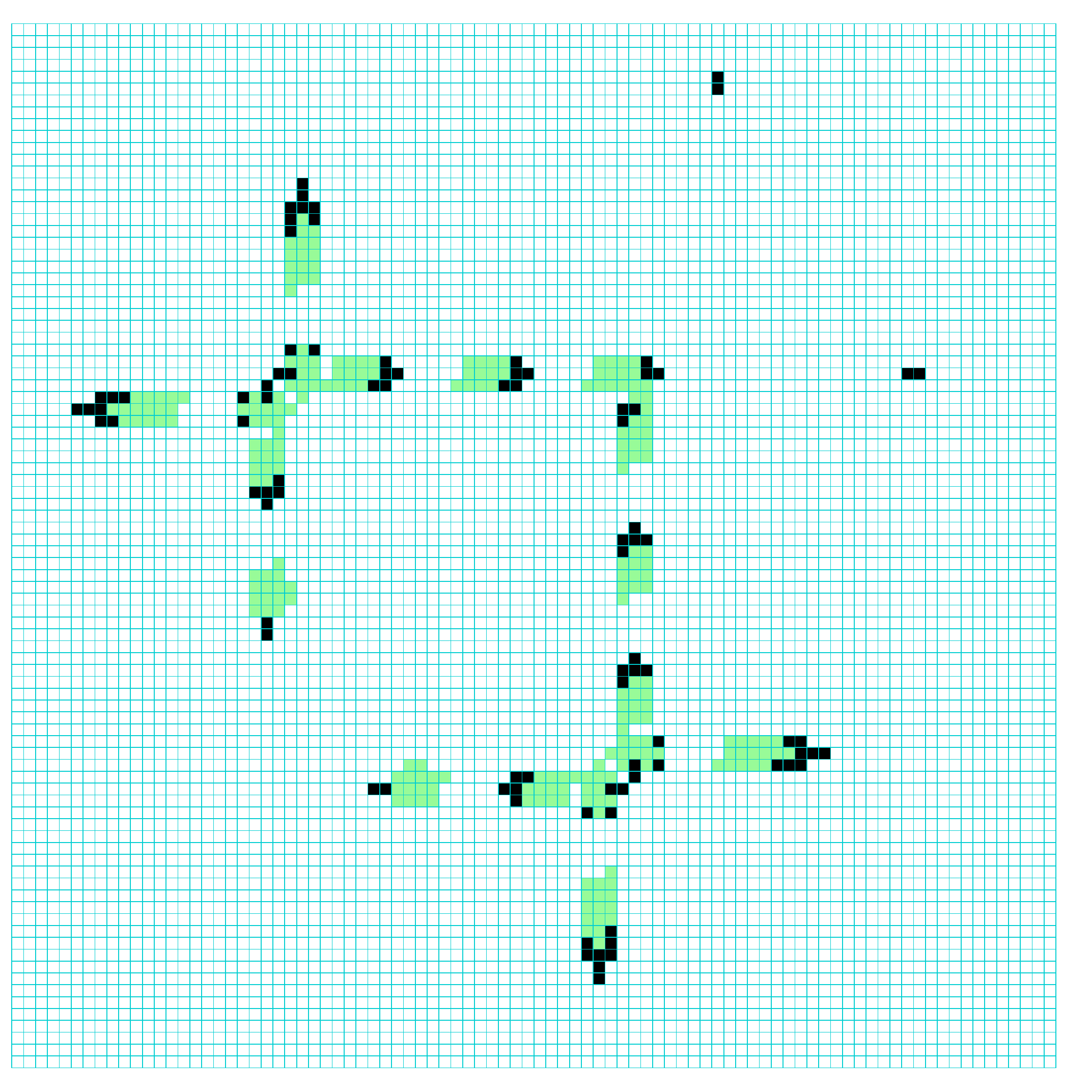} \hfill 
\includegraphics[width=.13\linewidth,bb=322 342 399 448, clip=]{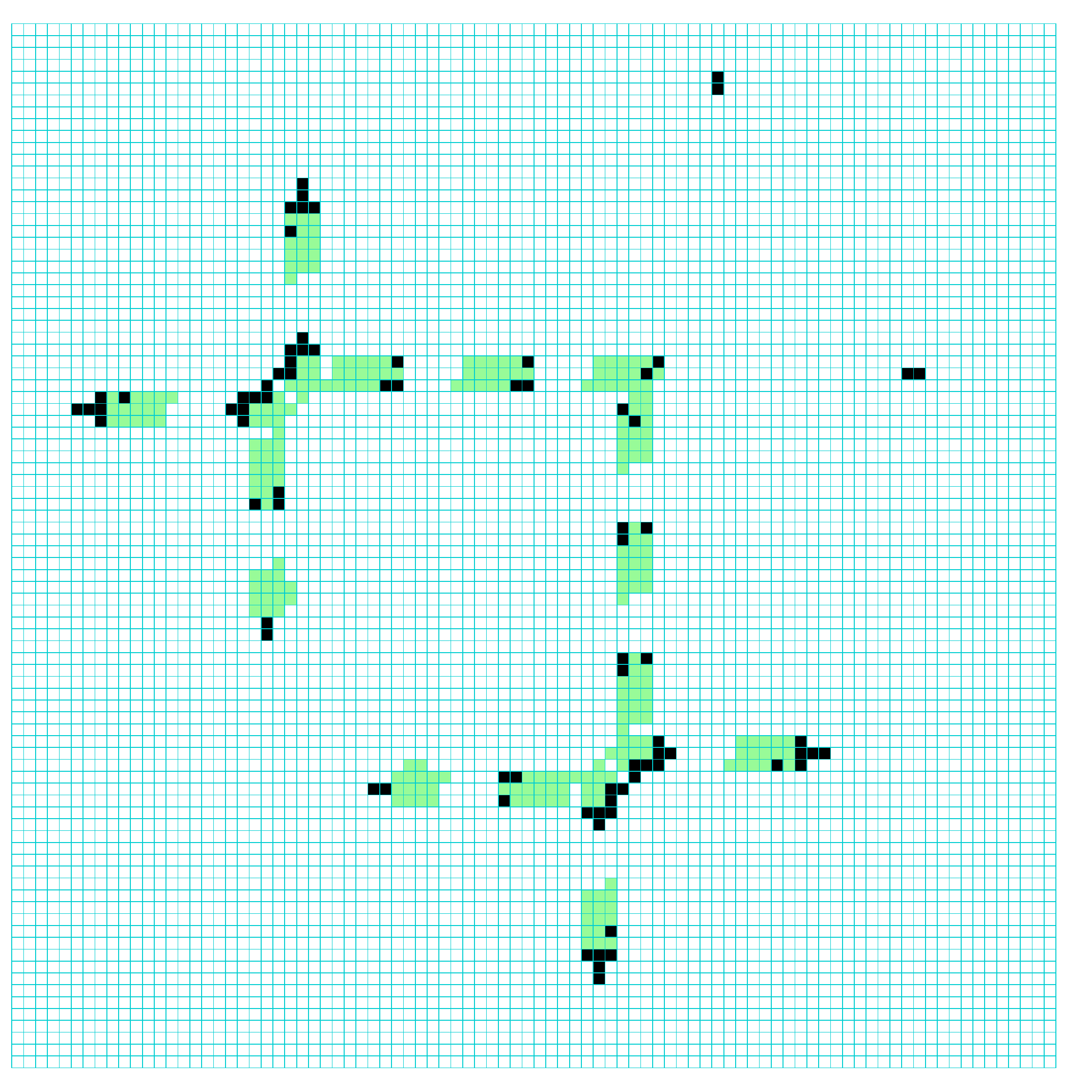} \hfill 
\includegraphics[width=.13\linewidth,bb=322 342 399 448, clip=]{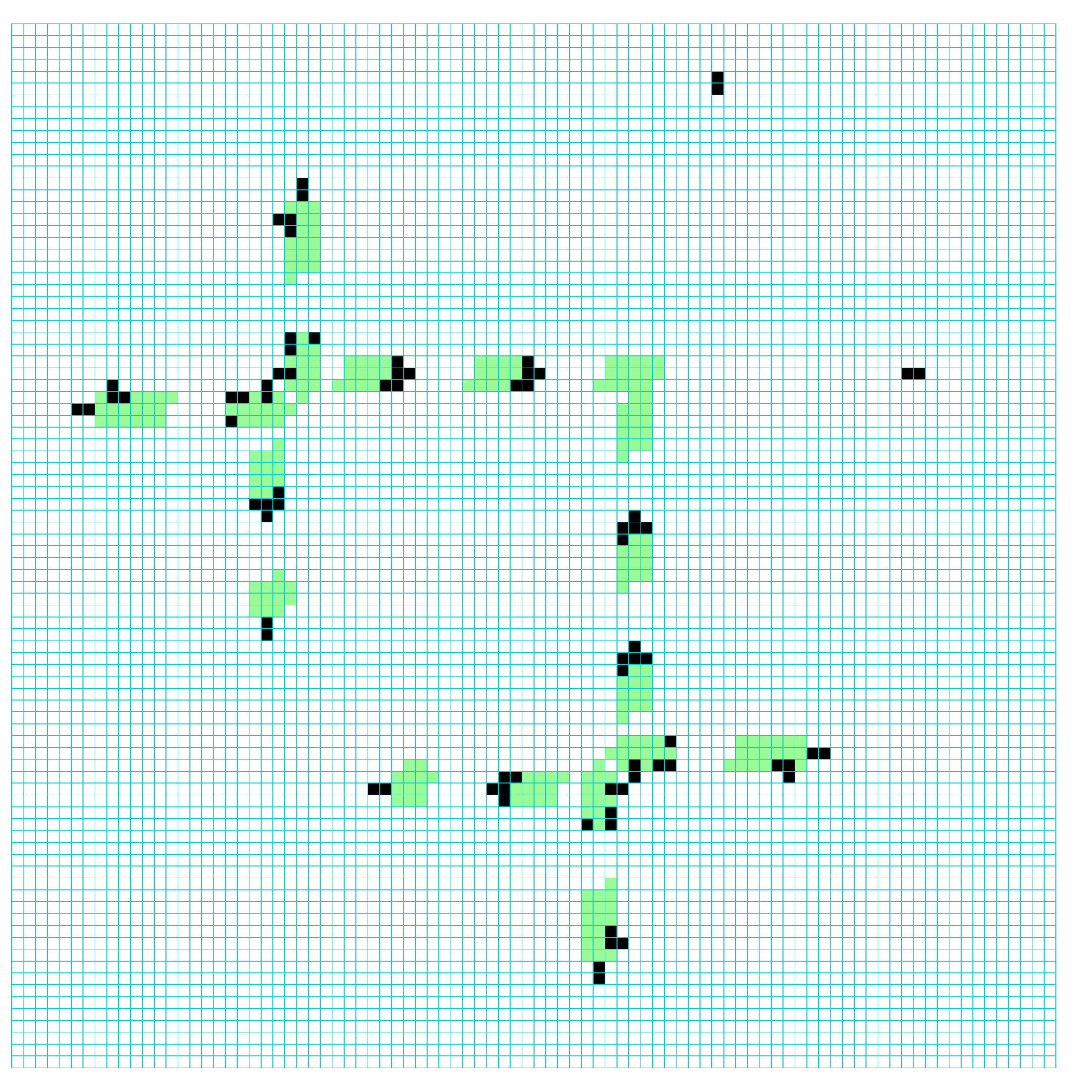} 
\end{minipage}
\end{center} 
\vspace{-3ex}
\caption[sideways glider collision]
{\textsf{
A sideways glider collisions between two gliders approaching at
90$^{\circ}$ can be arranged so that the gliders
self-destruct leaving no residue. The resulting gap in the
glider-gun-stream is wide enough for the following glider to pass
through. Green trails indicate motion.
\label{sideway glider collisions}
}}
\end{figure}

\enlargethispage{4ex}
Another essential requirement is that gliders colliding sideways can
self-destruct leaving no residue, and that the resulting gap in the
glider-gun-stream is sufficiently wide to allow a following glider to pass
through. The Ameyalli-rule satisfies all these requirements as illustrated in
figures~\ref{glidergun} to \ref{fig OR gate}.
\clearpage

\section{Logical Universality}
\label{Logical Universality} 

Traditionally the proof for universality in CA is based on the
Turing Machine or an equivalent mechanism, but in another approach by
Conway\cite{Berlekamp1982}, a CA is
universal in the full sense if it is capable of the following,

\begin{s-enumerate}
\item Data storage or memory.
\item Data transmission requiring wires and an internal clock.
\item Data processing requiring a universal set of logic gates NOT, AND, and OR, to satisfy
  negation, conjunction and disjunction.
\end{s-enumerate} 

This paper is confined to proving condition 3 only --- for universality in the logical sense.
To demonstrate universality in the full sense as for the
Game-of-Life, it would be necessary to also prove conditions 1 and 2,
or to prove universality in terms of the Turing Machine, as was done
by Randall\cite{Randall2002} for the Game-of-Life.
\clearpage

\section{Logical Gates}
\label{Logical Gates}

Logical universality in the Ameyalli-rule, as in the Game-of-Life,
is based on Post's Functional Completeness Theorem (FCT)\cite{Francis90}. 
This theorem guarantees that it is
possible to construct a conjunctive (or disjunctive) normal form formula
using only the logical gates NOT, AND and OR.

\begin{figure}[htb]
\begin{center} 
\includegraphics[width=1\linewidth]{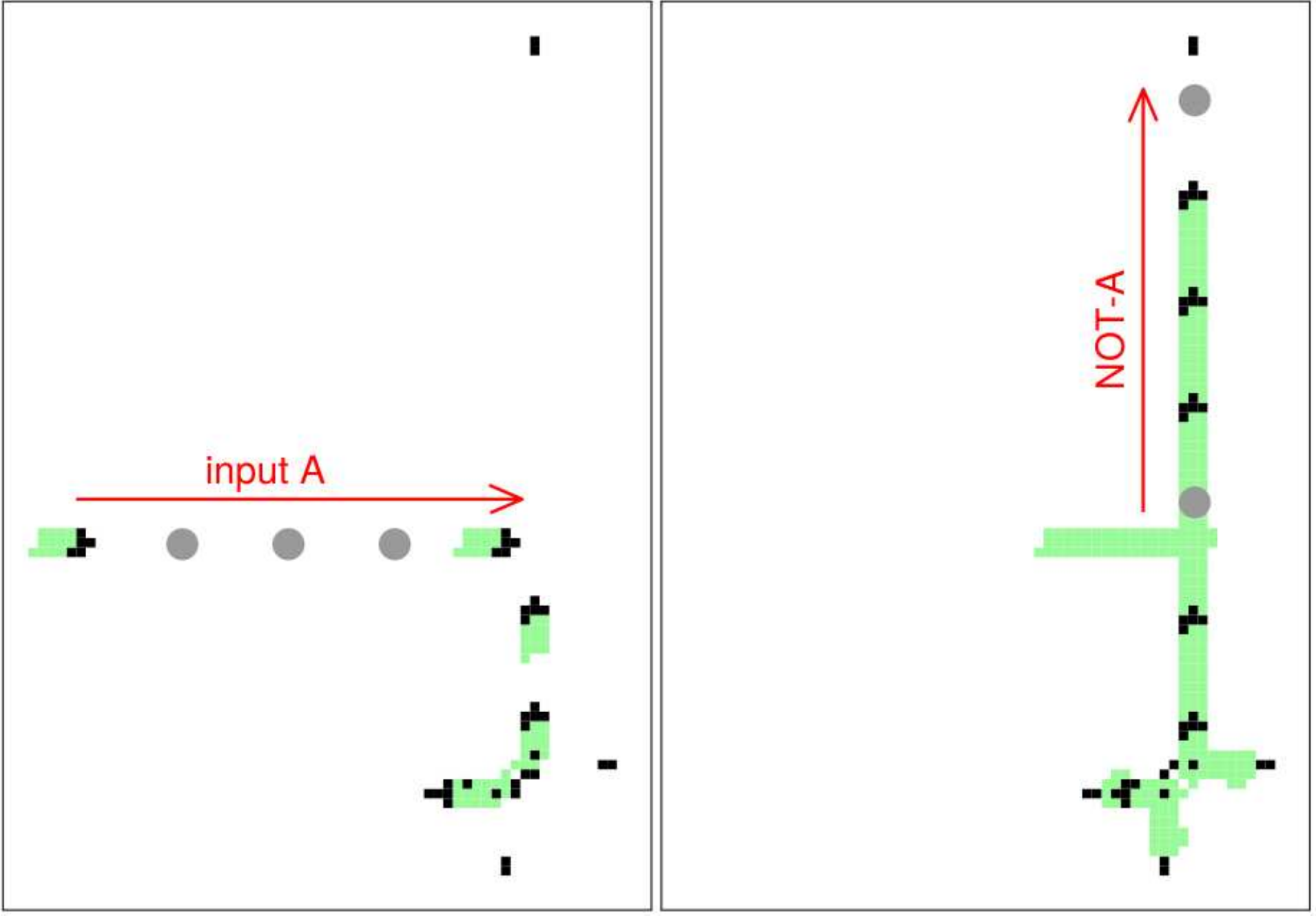}
\end{center}
\vspace{-3ex}
\caption[NOT gate]
{\textsf{
    An example of a NOT gate: ($\neg 1, 1 \rightarrow$ 0 and $\neg 0, 0 \rightarrow$ 1)
    or inverter, which transforms a stream of data to
    its complement, represented by gliders and gaps (grey discs). 
$Left$: The 5-bit input string A (10001) moving East is about to interact 
with a glider-stream moving North. $Right$: The outcome is NOT-A (01110) moving
North, shown after about 133 time-steps.
\label{fig NOT gate}
}}
\end{figure}

\begin{figure}[htb]
\begin{center} 
\begin{minipage}[c]{1\linewidth}
\fbox{\includegraphics[height=.68\linewidth,bb=40 6 301 358, clip=]{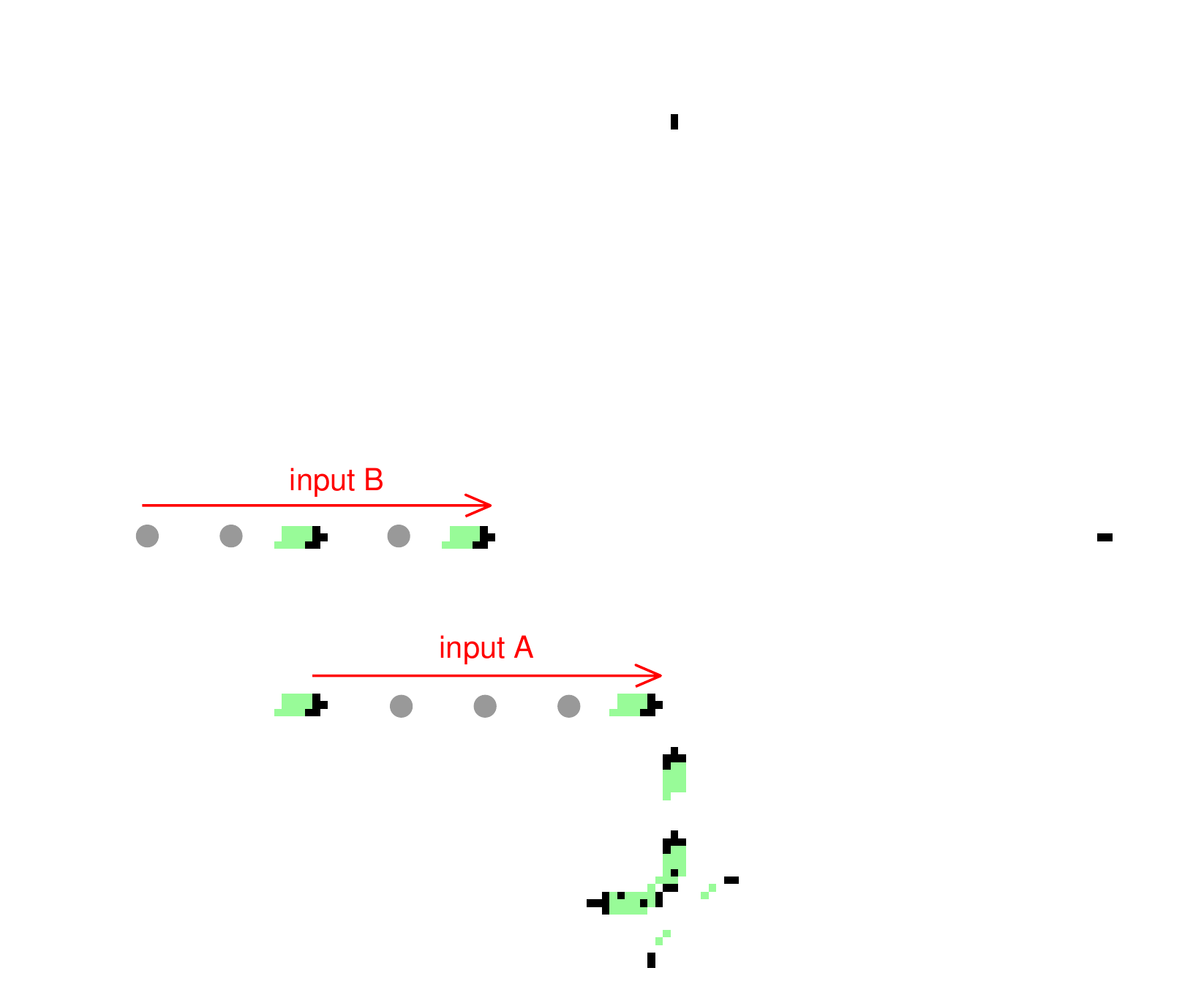}}
\hfill
\fbox{\includegraphics[height=.68\linewidth,bb=213 6 448 358, clip=]{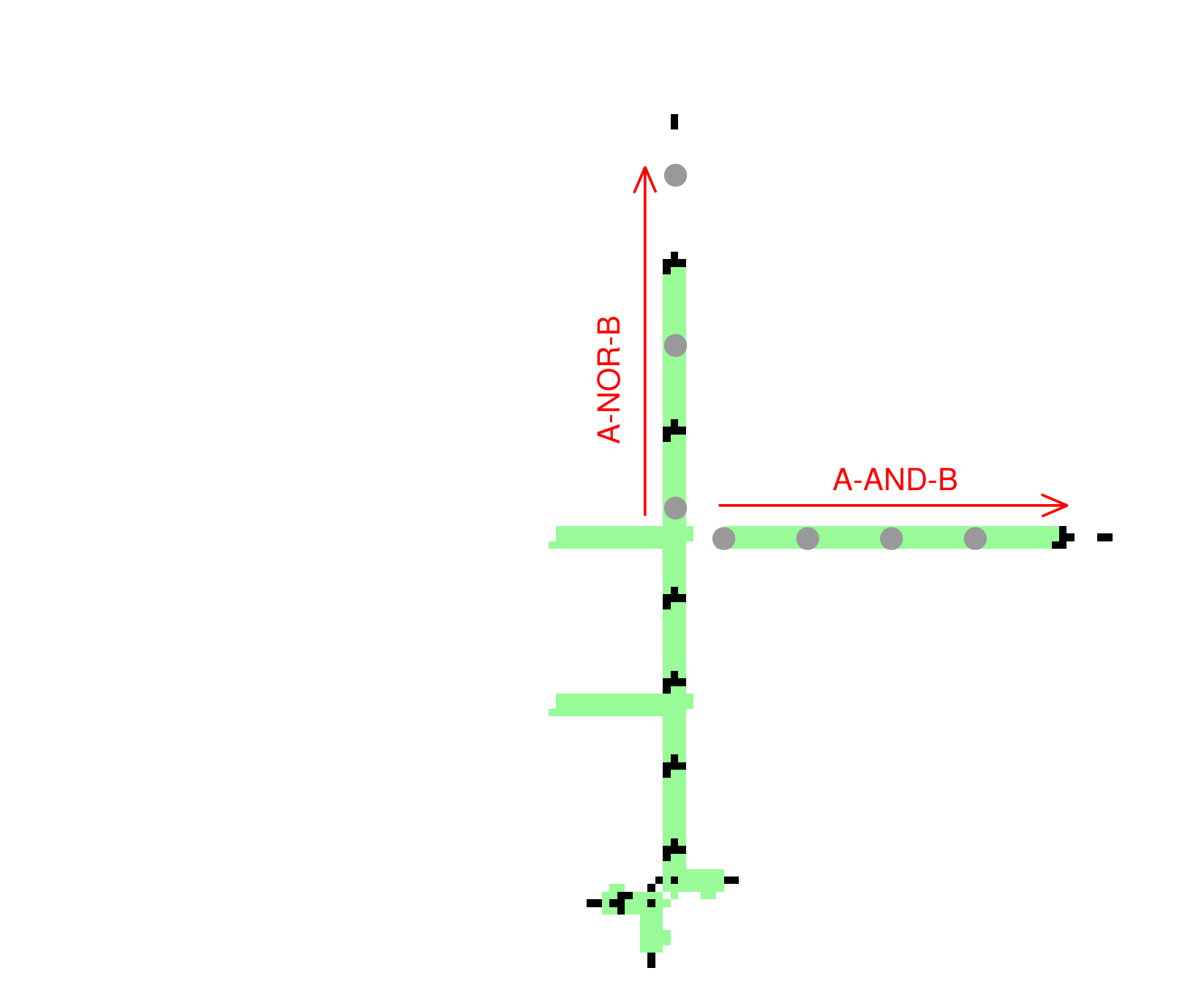}}
\end{minipage}
\end{center}
\vspace{-3ex}
\caption[AND gate]
{\textsf{
An example of the AND gate (1 $\wedge$ 1 $\rightarrow$ 1,
    else $\rightarrow$ 0) making a conjunction between two streams of
    data, represented by gliders and gaps (grey discs).
$Left$: The 5-bit input strings A (10001) and B (10100) both moving East
are about to interact with a glider-stream moving North. $Right$: The outcome is
\mbox{A-AND-B} (10000) moving East shown after about 197 time-steps.\\
{\color{white}xxx}
The dynamics making this AND gate first makes an intermediate NOT-A
(North~01110 -- figure~\ref{fig NOT gate}) which interacts with
input B to simultaneously produce both \mbox{A-AND-B} (East~10000),
and the A-NOR-B (North~01010) which will
be required to make the OR gate in figure~\ref{fig OR gate}.
\label{fig AND gate}
}}
\end{figure}

\begin{figure}[htb]
\begin{center} 
\begin{minipage}[c]{1\linewidth}
\fbox{\includegraphics[width=1\linewidth,bb=247 72 1093 526, clip=]{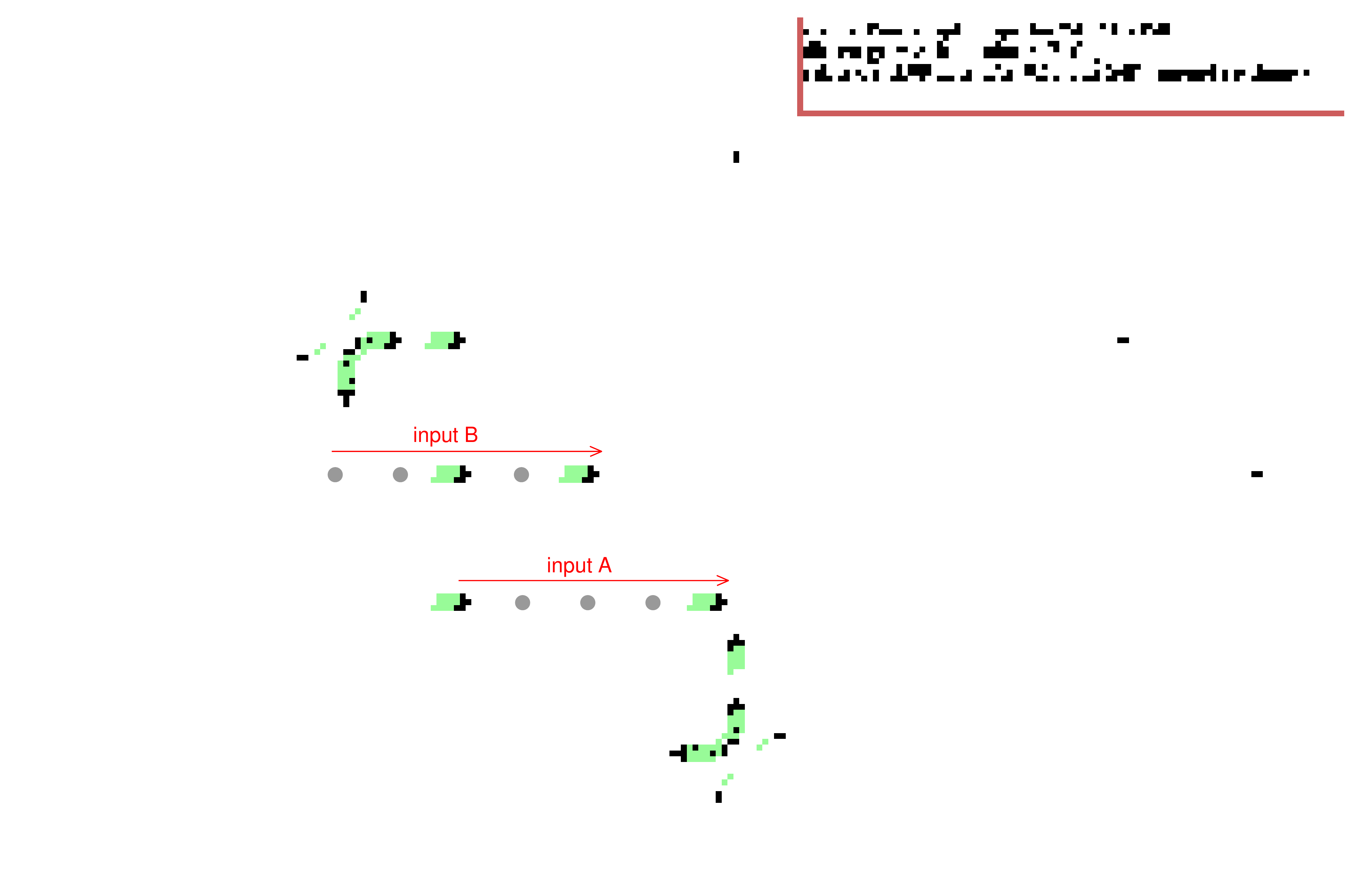}}\\[.5ex]
\fbox{\includegraphics[width=1\linewidth,bb=247 72 1098 526, clip=]{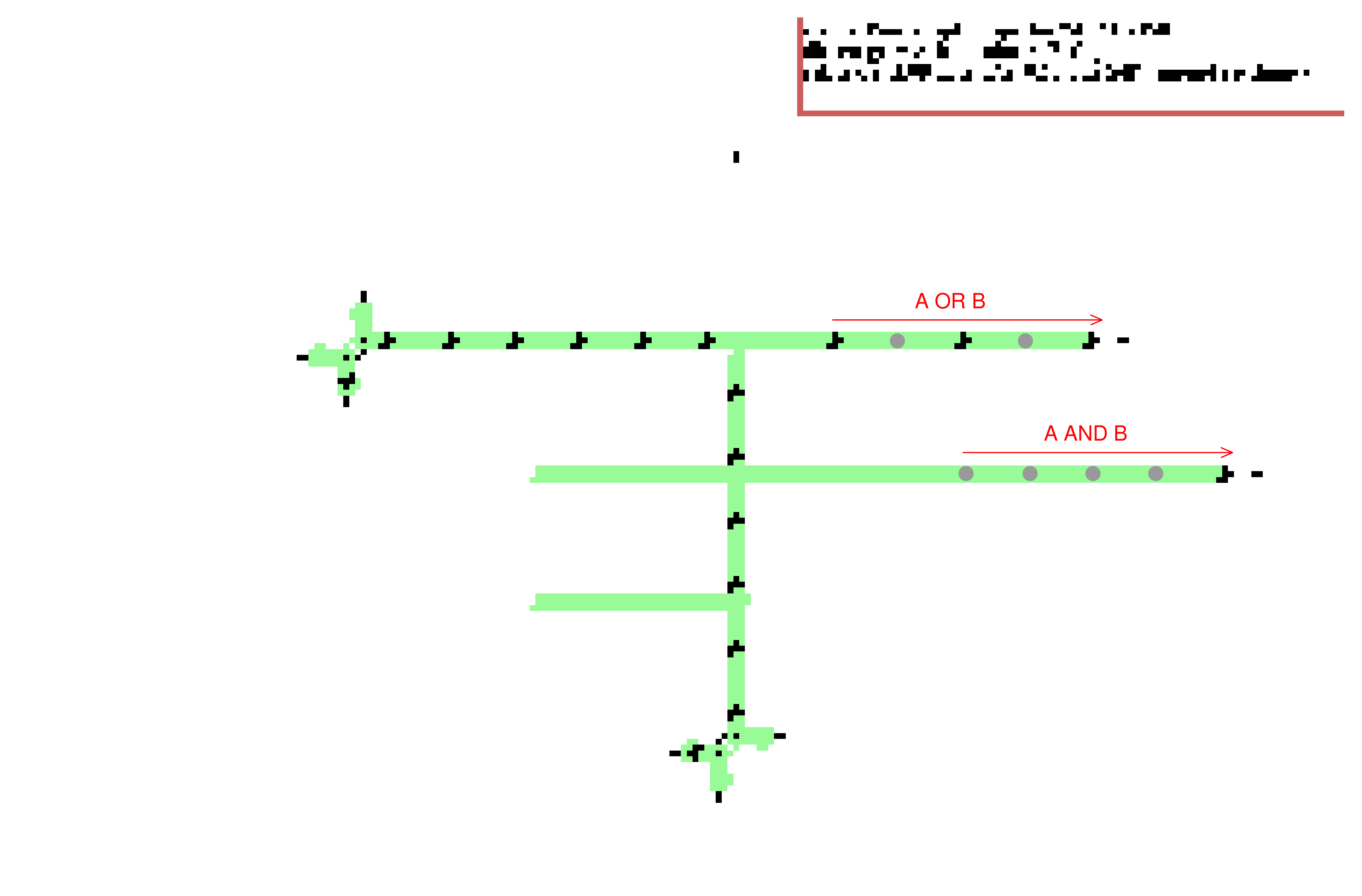}}
\end{minipage}
\end{center}
\vspace{-3ex}
\caption[OR gate]
{\textsf{
An example of the OR gate (1 $\vee$ 1 $\rightarrow$ 1, else
$\rightarrow$ 0) making a disjunction between two stream of data
represented by two streams of gliders and gaps (grey discs).
$Top$: The 5-bit input strings A (10001) and B (10100) both moving East
are about to interact with two glider-streams, the lower shooting North, and
the upper shooting East. $Below$: The outcome is A-OR-B
(10101) moving East shown after 220 time-steps.\\
{\color{white}xxx}
The dynamics making this OR gate first makes an intermediate NOT-A
(North~01110 -- figure~\ref{fig NOT gate}) which interacts with
input B to make A-NOR-B (North~01010 -- figure \ref{fig AND gate})
which interacts with
the upper glider-stream shooting East to make A-OR-B (East~10101).
A residual bi-product is A-AND-B (East~10000 -- figure \ref{fig AND gate}).
\label{fig OR gate}
}}
\end{figure}

Using a specific right-angle collision, two gliders can self-destruct
leaving no residue as shown in figure~\ref{sideway glider collisions}. 
Applying this property between glider-gun streams
and a glider/gap sequence with the correct
spacing and phases representing a ``string'' of information, 
its possible to build the logical gates NOT, AND and OR,
illustrated in figures~\ref{fig NOT gate}, \ref{fig AND gate} and \ref{fig OR gate}. 
Gaps in a string are indicated by grey circles, dynamic trails
are included, and B-type eaters are positioned to eventually
stop gliders. Note that the leading bit of
a moving glider/gap information sequence appears on the left
in its string representation.

\subsection{NOT gate}
\label{NOT gate} 

\clearpage

\subsection{AND gate}
\label{AND gate}

\clearpage

\subsection{OR gate}
\label{OR gate}

\enlargethispage{3ex}

\clearpage
\section{The Ameyalli-Rule definition}
\label{The Ameyalli-Rule definition}

\begin{figure}[htb]
{\textsf{\small
\begin{minipage}[c]{1\linewidth}
\includegraphics[width=1\linewidth]{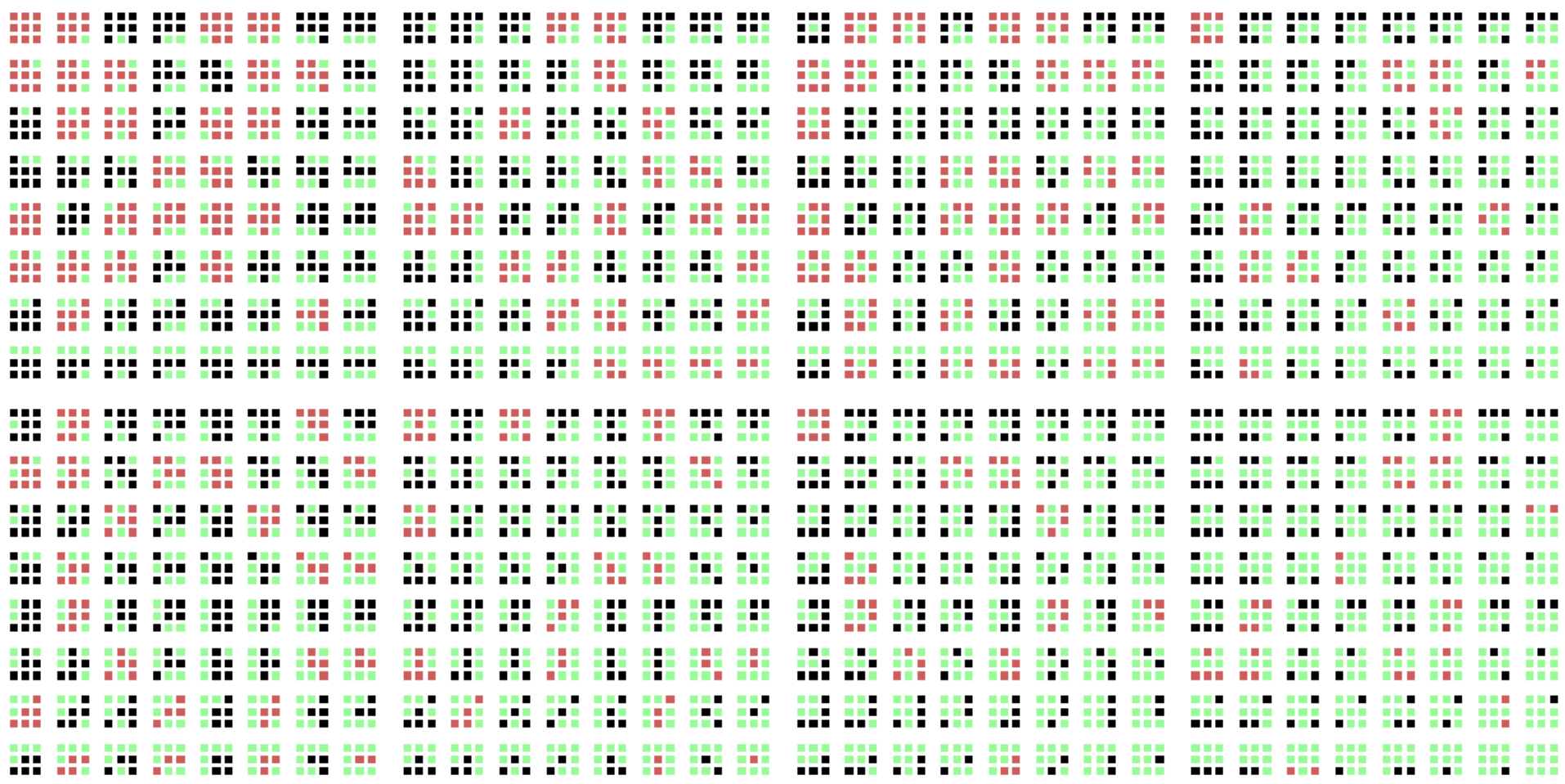}\\[1ex]
(a) Neighborhood patterns in descending decimal equivalent order from 
\raisebox{-.8ex}{\includegraphics[height=3ex]{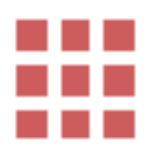}} -
to
\raisebox{-.8ex}{\includegraphics[height=3ex]{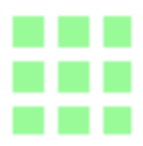}}.
(511 to 0) if each row from the upper matrix is continued in the lower matrix.
Ameyalli-rule neighborhoods that input 1 are colored red.
\end{minipage}\\[2ex]
\begin{minipage}[c]{1\linewidth}
  \includegraphics[width=1\linewidth,bb=3 12 549 81, clip=]{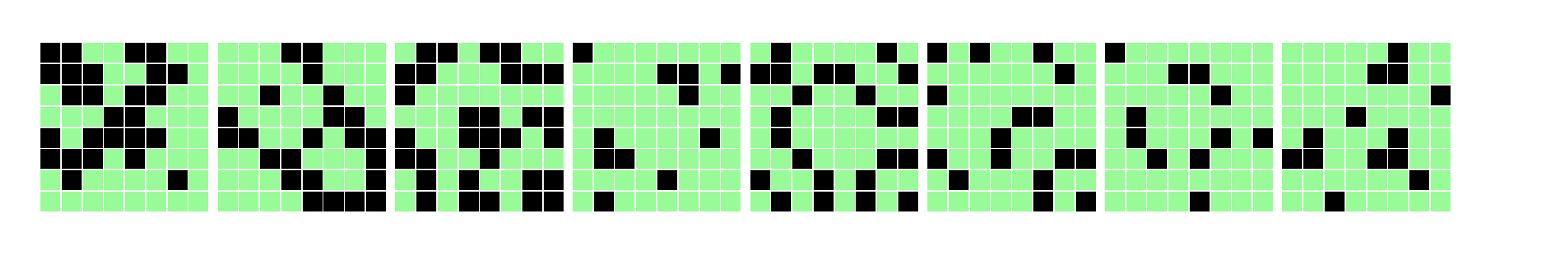}\\[1ex]
(b) The Ameyalli-rule showing just inputs follows the upper and lower matrices above,
set alongside each other giving a continuous descending decimal equivalent order 511 to 0.
This is the full rule-table (rcode). 154 inputs=1 are colored black.
$\lambda$=154/512=0.3.
\end{minipage}
}}
\vspace{-2ex}
\caption[The rcode showing nhoods]
{\textsf{(a) The Ameyalli-rule showing details of all neighborhood patterns,
and (b) the full rule-table, rcode in DDLab.
To restrict the table to isotropic rules only, algorithms ensure that all related
neighborhoods by spins/flips give the same input.
In these presentations the diagonal symmetry of each 8$\times$8 block
is a necessary but insufficient indication of isotropy.
\label{am-rall}
}}
\end{figure}

An isotropic CA rule based on a 2d binary 3$\times$3 Moore neighborhood ---
\raisebox{-.8ex}
{\includegraphics[height=3ex, bb=10 14 37 40,clip=]{am_pdf-figs/am-gg22}}
--- can be defined by a series of methods that become ever simpler, clearer, and more concise,
illustrated in figures~\ref{am-rall}, \ref{am101-0}, and \ref{am102}.
In all these methods, a descending order (from left to right) of the neighborhood's
decimal equivalent is employed, in line with Wolfram's classic
convention\cite{wolfram83,wolfram2002} --- rule-tables can then be expressed in decimal
or hexadecimal\cite{EDD}.

The decimal equivalent of a 3$\times3$
pattern is taken as a string in the order
\begin{minipage}[c]{4ex}\scriptsize 876\\[-1ex]543\\[-1ex]210\end{minipage} --- for example,  
the pattern 
\raisebox{-.8ex}{\includegraphics[height=3ex]{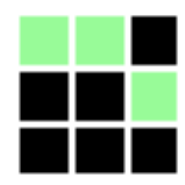}} 
is the binary string 001110111 (119 in decimal), 
representing the full rule-table index 119 (figure~\ref{am-rall}),
the symmetry class 119 (figure~\ref{am101-0}),
and iso-rule index 66 (figure \ref{am102}). 

\subsection{full lookup-table}
\label{full lookup-table}
\vspace{-2ex}

\enlargethispage{1ex}

\clearpage
\subsection{symmetry class table --- iso-groups}
\label{symmetry class table --- iso-groups}
\vspace{-2ex}
  
\enlargethispage{2ex}
\begin{figure}[htb]
  \includegraphics[width=1\linewidth]{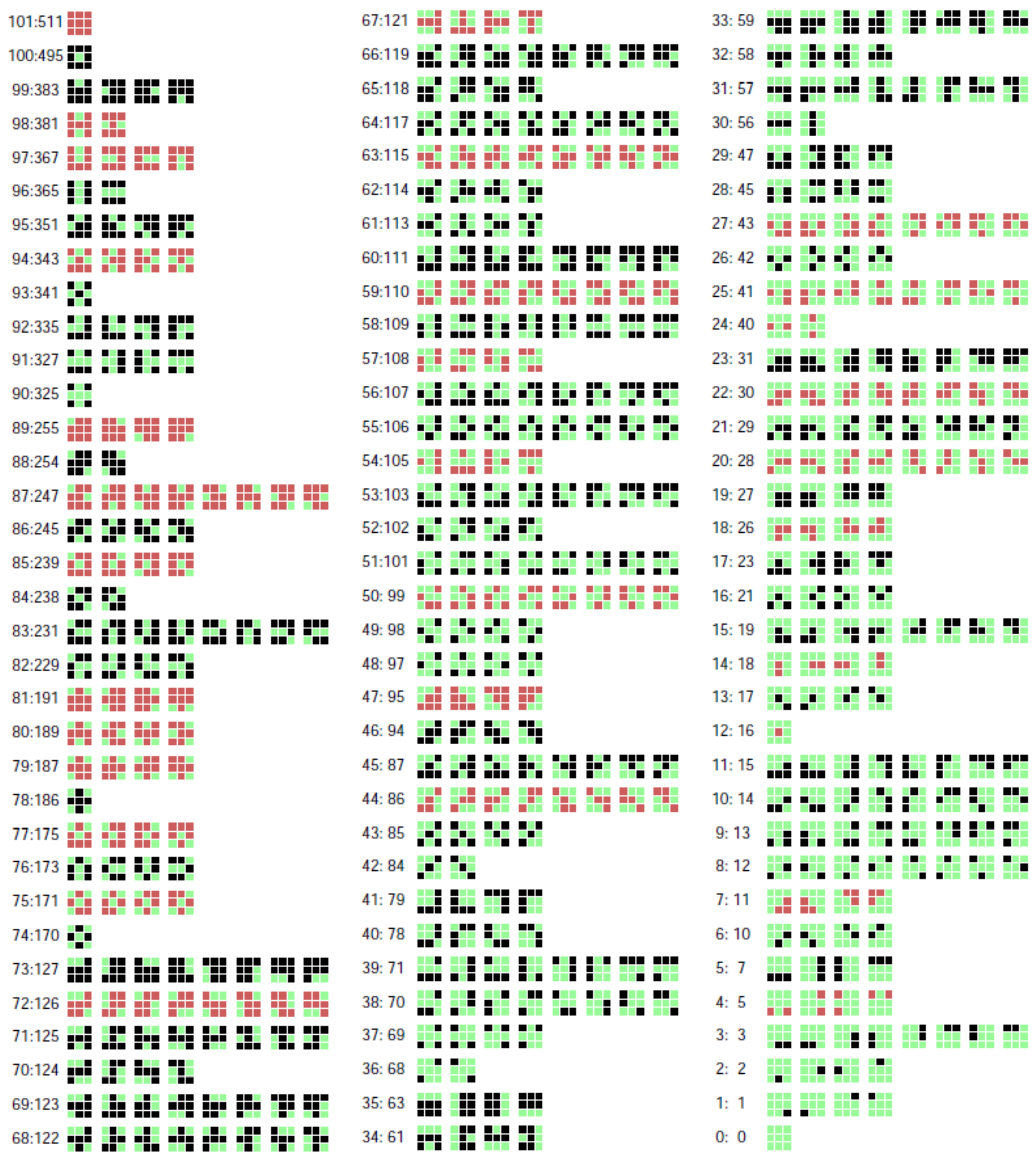}\\
 \textsf{\small  \phantom{xxxxx}101-68  \phantom{xxxxxxxxxxxxxxxxxxxxx}67-34 \phantom{xxxxxxxxxxxxxxxxxxxxx}33-0}
\vspace{-2ex}
\caption[symmetry class table isotropic]
{\textsf{
The symmetry class table (equivalent to iso-groups in figure\ref{am102})
lists all 102 prototype neighborhoods together with their group of spins/flips 
listed in sequence --- here in 3 columns. The prototype is the smallest decimal
equivalent and serves as the symmetry class index which is
discontinuous between 511 and 0. Patterns within each symmetry class are ordered from 
lowest (left) to highest decimal equivalent --- opposite to the usual order.
This convention was employed in \cite{Gomez2015,Gomez2017,Gomez2018,Gomez2020} where
only symmetry classes with inputs 1 are usually listed, but here we show
all 102. Input 1 neighborhoods are colored red.
The Ameyalli-Rule is defined by the 31 red symmetry classes
which input 1, the rest input 0. 
Numbered labels (101:511 to 0:0) show the iso-rule index 101 to 0, 
followed by the symmetry class index.
\label{am101-0}
}}
\end{figure}

\subsection{iso-rule}
\label{iso-rule}

\begin{figure}[htb]
  \begin{center}
  \begin{minipage}[t]{.8\linewidth}
    \includegraphics[width=1\linewidth,bb=50 705 630 890, clip=]{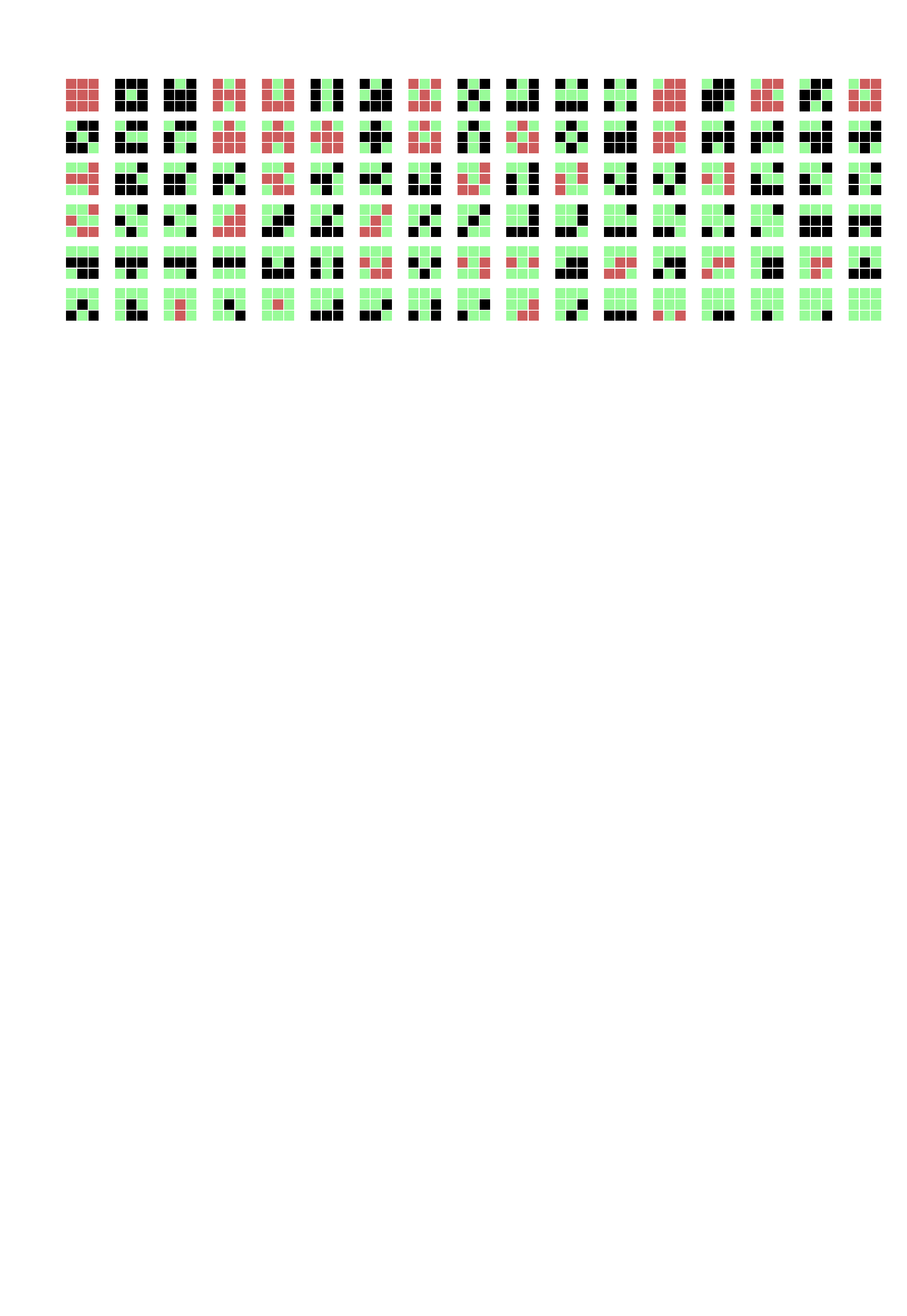}\\[-.5ex] 
  \textsf{\small (a) 102 iso-group prototypes}
  \end{minipage}\\[2ex]
  \end{center}
\begin{minipage}[t]{1\linewidth}
  \includegraphics[width=1\linewidth,bb=8 13 824 25, clip=]{am_pdf-figs/am-iso}\\[-.2ex] 
  \textsf{\small (b) the iso-rule indexed 101-0, left to right.}
\end{minipage}
\vspace{-1ex}
\caption[The iso-rule]
{\textsf{(a) The 102 iso-group prototypes indexed 101 to 0.
The prototype order is from highest decimal equivalent (511 top-left) to lowest (0 bottom-right),
which is also the left to right order of as iso-rule, indexed 101 to 0, the simplest expression 
of an isotropic rule.
The Ameyalli is defined by the 31 red prototypes 
which input 1, the rest input 0.\\
(b) The Ameyalli iso-rule compacts the information in (b) into a simple bit-string,
here shown as a DDLab graphic with 1/0 inputs shown as black/green.\\
The iso-rule is expressed in hexadecimal (iso-hex)=2642a3a9088a4490000b545090
by breaking the bit-string into 4-bit segments (with a complete segment on the right).
For example, the hexadecimals of the 7 segments on the right $\dots$b545090 are given by
$\dots$1011=b, 0101=5, 0100=4, 0101=5 ,0000=0, 1001=9, 0000=0.\\
The Hensel string translated for Golly\cite{Golly}, 
generated automatically in DDLab\cite{Wuensche-DDLab,EDD} is
B2ci3ar4krtz5cq6c7ce/S01e2ek3qj4kt5ceayq6cki7c8.
\label{am102}
}}
\end{figure}

The iso-rule\cite{Wuensche2021,EDD} depends on just the 102 prototype
neighborhoods in figure~\ref{am102}(a) with a rule-table index 101-0,
following the symmetry class table in figure~\ref{am101-0} but
with the rest of the group taken as read and omitted.  The iso-rule
shown in figure~\ref{am102}(b) is then just a 102 length bit-string 
(indexed 101-0) listing prototype inputs, 1 or 0.

Besides simplicity and brevity, an advantage of the iso-rule is that any
mutation conserves isotropy. The iso-rule 
has many other advantages\footnote{The iso-rule method applies in general to
a variety of CA neighborhood templates, in 1d, hex/square-2d,
and 3d, including multi-value as well as binary\cite{Wuensche2021,EDD}.}\cite{Wuensche2021,EDD} 
when studying isotropic CA as compared to a full or symmetry class
rule-table. The iso-rule provides a more direct entropy scatter plot for automatically
classifying isotropic rule-space, and as shown in
section~\ref{iso-rule input-frequency-histogram (IFH)}, it provides a
concise input-frequency histogram (IFH) with its functions of mutation and
filtering.

\section{iso-rule input-frequency-histogram (IFH)}
\label{iso-rule input-frequency-histogram (IFH)}

The input frequency histogram (IFH)\cite{Wuensche99,Wuensche2021},
a method in DDLab\cite{Wuensche-DDLab,EDD}, allows a dynamic view of the activity
of a rule-table in relation to the current iterating space-time pattern,
with options to interactively filter and mutate the current rule. Any isolated periodic pattern
can be investigated to ascertain which rule-table inputs\footnote{The IFH applies to
any type of rule-table, full and totalistic as well as iso-rules.}
are responsible for maintaining the pattern and to what extent.

\begin{figure}[htb]
\textsf{\small
\includegraphics[height=.16\linewidth,bb=32 65 200 209, clip=]{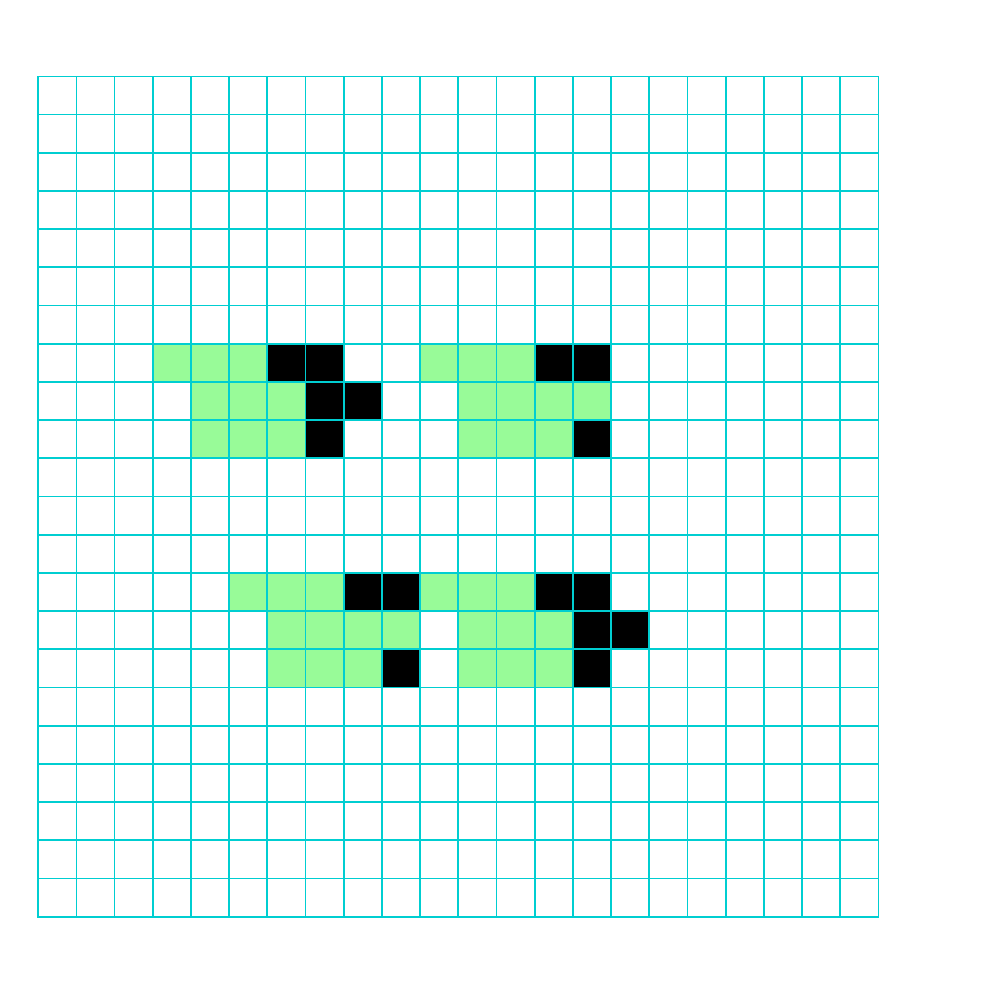}
\includegraphics[height=.16\linewidth,bb=32 65 200 209, clip=]{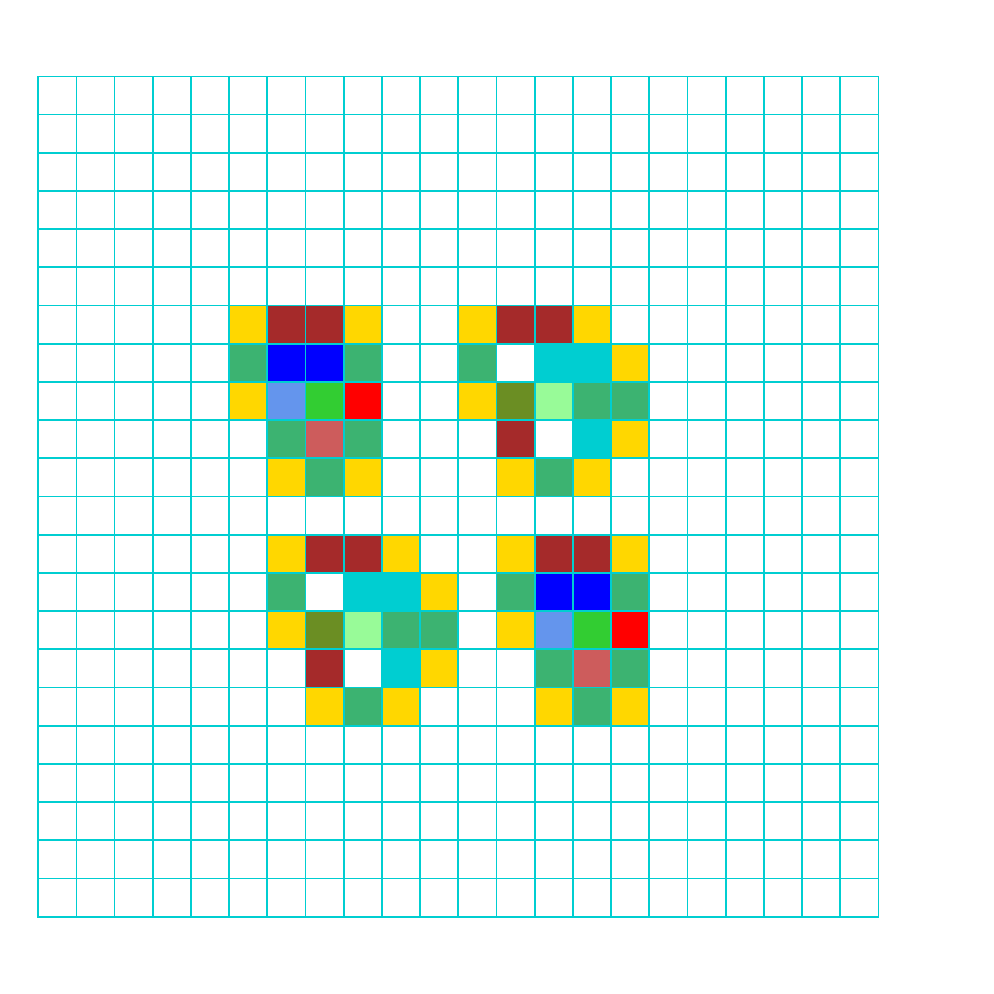}
\phantom{i}
\raisebox{9ex}{
  \begin{minipage}[t]{.55\linewidth}
{\footnotesize Four Ameyalli gliders moving East.\\
{\it far left}: colors by value with green \mbox{dynamic trails.}\\
{\it near left}: colors corresponding to IFH colors.}
  \end{minipage}
}\\
\begin{minipage}[c]{1\linewidth}
\includegraphics[width=1\linewidth,height=.25\linewidth]{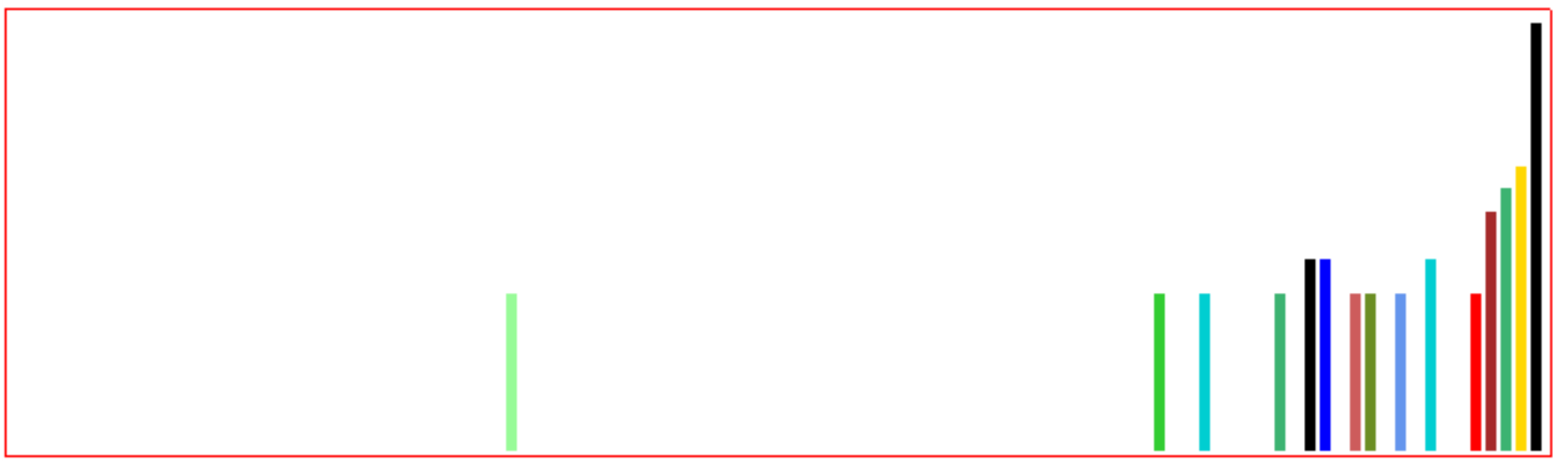}
\end{minipage}
}
\vspace{-2ex}
\caption[Ameyalli-rule glider IFH]{
  \textsf{The IFH for the Ameyalli glider, with 15 active bars.
}}
\label{glider IFH}
\end{figure}

\begin{figure}[htb]
\textsf{\small
\setlength{\fboxsep}{0pt}\fbox{\includegraphics[height=.18\linewidth]{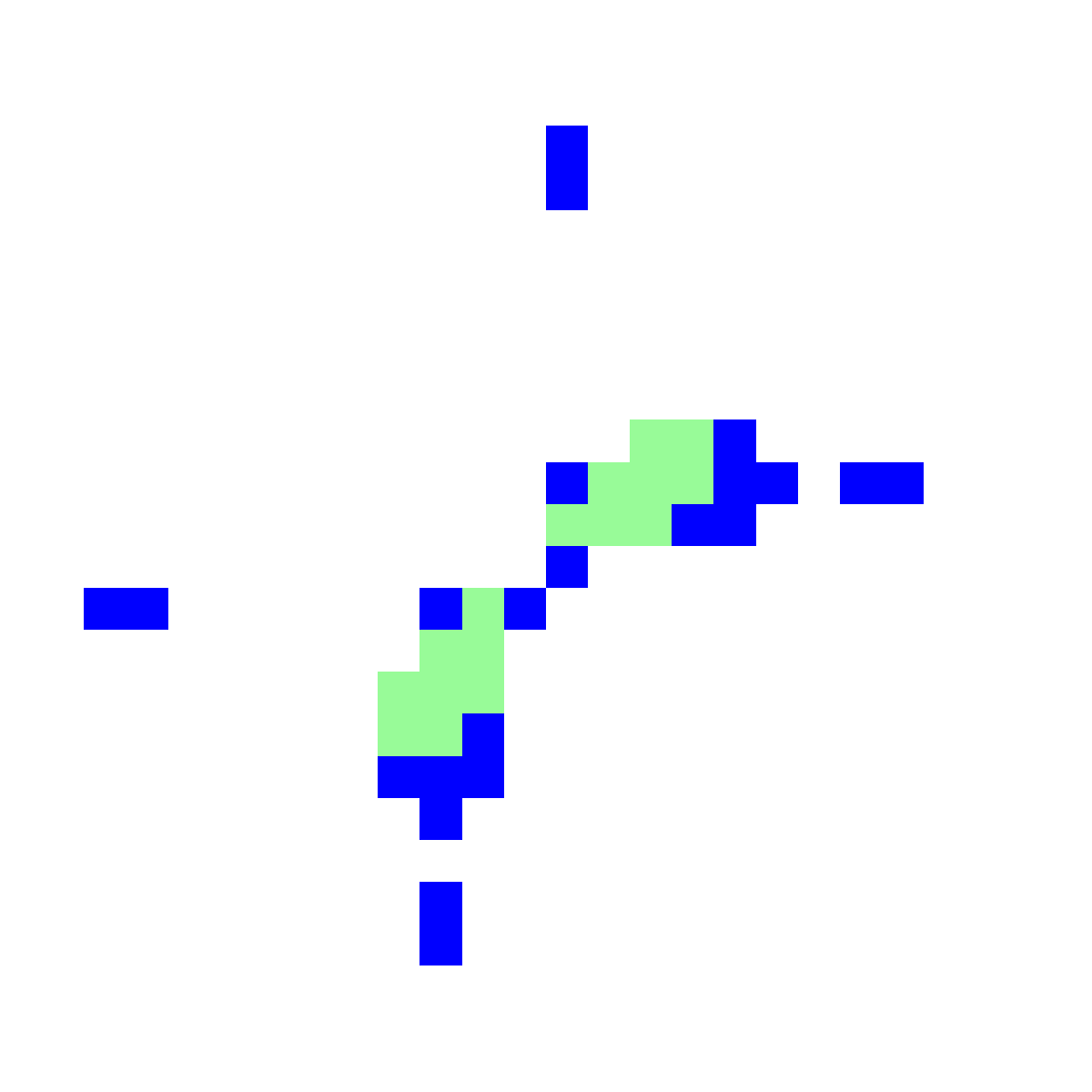}}
\setlength{\fboxsep}{0pt}\fbox{\includegraphics[height=.18\linewidth]{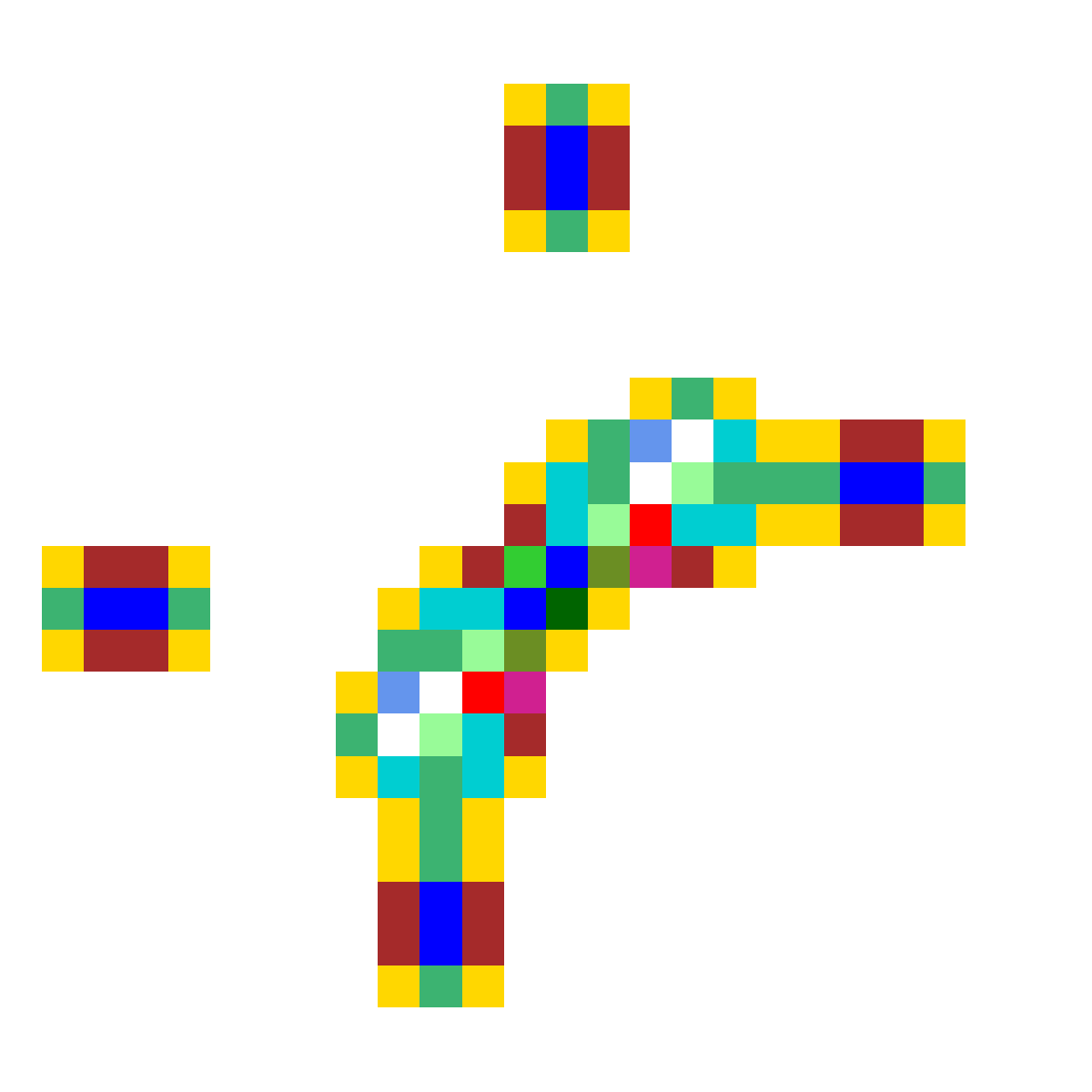}}
\phantom{i}
\raisebox{11ex}{
  \begin{minipage}[t]{.55\linewidth}
 {\footnotesize The Ameyalli-rule glider-gun contained by type-B eaters as in figure~\ref{attractor-cycle}.\\
{\it far left}: colors by value with green \mbox{dynamic trails.}\\
{\it near left}: colors corresponding to IFH colors.}
  \end{minipage}
}\\
\begin{minipage}[c]{1\linewidth}
\includegraphics[width=1\linewidth,height=.25\linewidth]{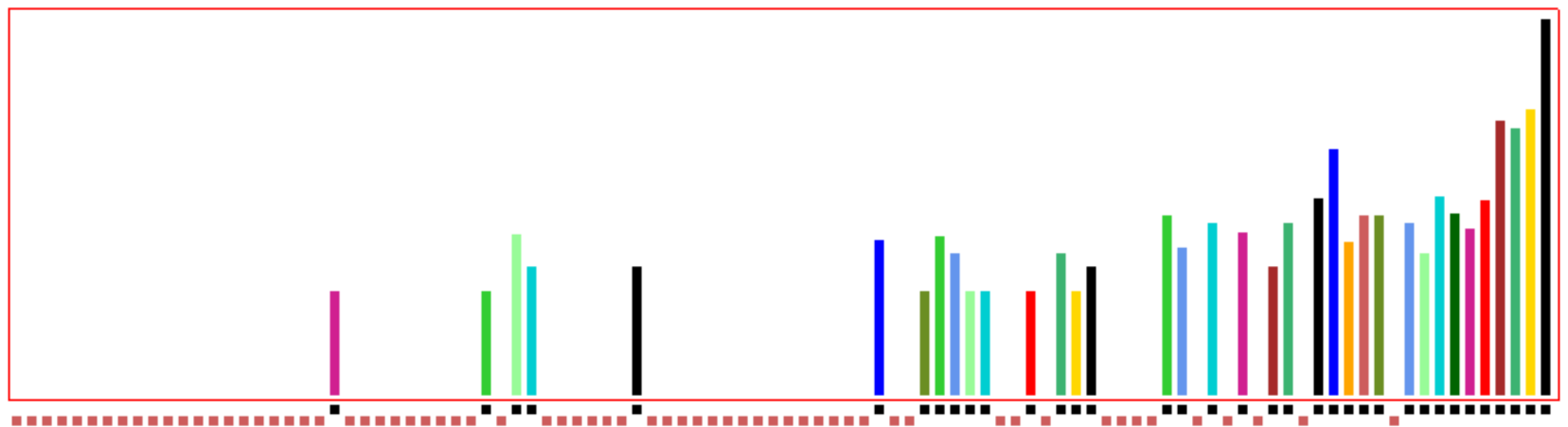}\\
\includegraphics[width=1\linewidth,bb=6 13 835 25, clip=]{am_pdf-figs/am-iso}\\[-.5ex] 
\includegraphics[width=1\linewidth,bb=6 13 835 25, clip=]{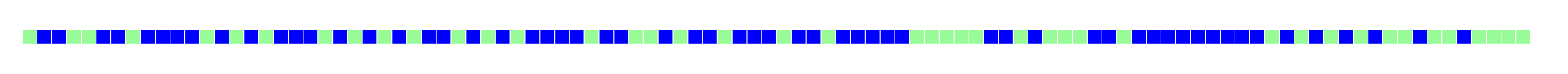}\\[-.5ex] 
\includegraphics[width=1\linewidth,bb=6 13 835 25, clip=]{am_pdf-figs/am-muts}\\[-.5ex] 
\textsf{\small \phantom{xxxxxxxxxxxxxxxxxxx}eater-B\phantom{xxxxxxxxxxxxxxxxxxxx}eater-A}
\end{minipage}
}
\vspace{-2ex}
\caption[Ameyalli-rule glider-gun IFH]{
  \textsf{The log$_2$ iso-rule IFH of the Ameyalli glider-gun/eater-B system.
  All 32 active bars were firstly filtered (black blocks), then all 70 neutral inputs were mutated
  (red blocks) --- the dynamics were not affected.
  Below the IFH the corresponding iso-rule tables are shown, firstly the original Ameyalli-rule, and
  below that the mutated rule. The position of entries for eater A and B are indicated, noting that there
  is no activity for the absent eater A.\\
The iso-rules in hexadecimal hex are compared below:}\\
\texttt{\small
  26 42 a3 a9 08 8a 44 90 00 0b 54 50 90} \textsf{---Ameyalli-rule}\\
\texttt{\small
  19 bd 5d 56 af 65 bb 7c 1a 37 fd 54 90} \textsf{---after all 70  inactive mutations}
}
\label{glider-gun IFH}
\end{figure}

For example, in figure~\ref{glider IFH} the space-time pattern
consists of four non-stop Ameyalli gliders on a lattice with periodic
boundaries. The IFH accommodates 102 columns to indicate the level of
activity\footnote{In figures~\ref{glider IFH},\ref{glider-gun IFH},\ref{idealized iso-rule IFH},
  the measures are averaged over a moving window of 100 time-steps to
  stabilise the IFH.  The number of trailing hits $h_i$ (lookups at index $i$
  of the iso-rule-table) at each time-step is recorded, and  converted to
  $f_i$, the fraction (0 to 1) of all possible hits
  $h_i$$/$$(n\times w\times S)$, where $n$ is lattice size,  $w$ (100) is the size
  of the  moving window of time-steps, and $S$~(102) is the iso-rule size.
  $f_i$ is converted to log$_2$ to amplify infrequent hits, which sets the column height.}
of each iso-group --- the frequency of iso-rule inputs, indexed from
101 to 0 (left to right) as in figure~\ref{am102}.  Only 15 bars
appear showing which inputs are active and to what extent, noting that
the bar height is log$_2$ of the actual frequency to amplify
infrequent hits.  Mutating the iso-rule at any of these bar indices
disrupts the gliders, but mutating at any other index makes no
difference to the gliders themselves, but would very probably disrupt
any other Ameyalli pattern.  We call such inactive inputs ``neutral''
relative to the particular space-time pattern.

In figure~\ref{glider-gun IFH} the (periodic) space-time pattern is
the Ameyalli glider-gun system contained by eaters-B, similar to
figure~\ref{attractor-cycle}, In this case 32 iso-groups are active,
the rest are neutral indicated by missing columns.  In DDLab,
on-the-fly key-hits will progressively filter active columns from high
to low, marked by black blocks.  Filtering excludes drawing these
cells in the space-time pattern, or alternatively just marks the
neutral bars.  Other key-hits progressively mutate the iso-rule at
column positions from low to high, marked by black blocks, so neutral
inputs are preferentially mutated, either by flipping the input, or
setting zero.  In this way its possible to mutate any or all neutral
inputs, and as expected, experiment confirms there is no effect on the
given glider-gun system.

\subsection{The idealized iso-rule}
\label{The idealized iso-rule}

\begin{figure}[htb]
\textsf{\small
  \setlength{\fboxsep}{0pt}\fbox{\includegraphics[height=.25\linewidth]{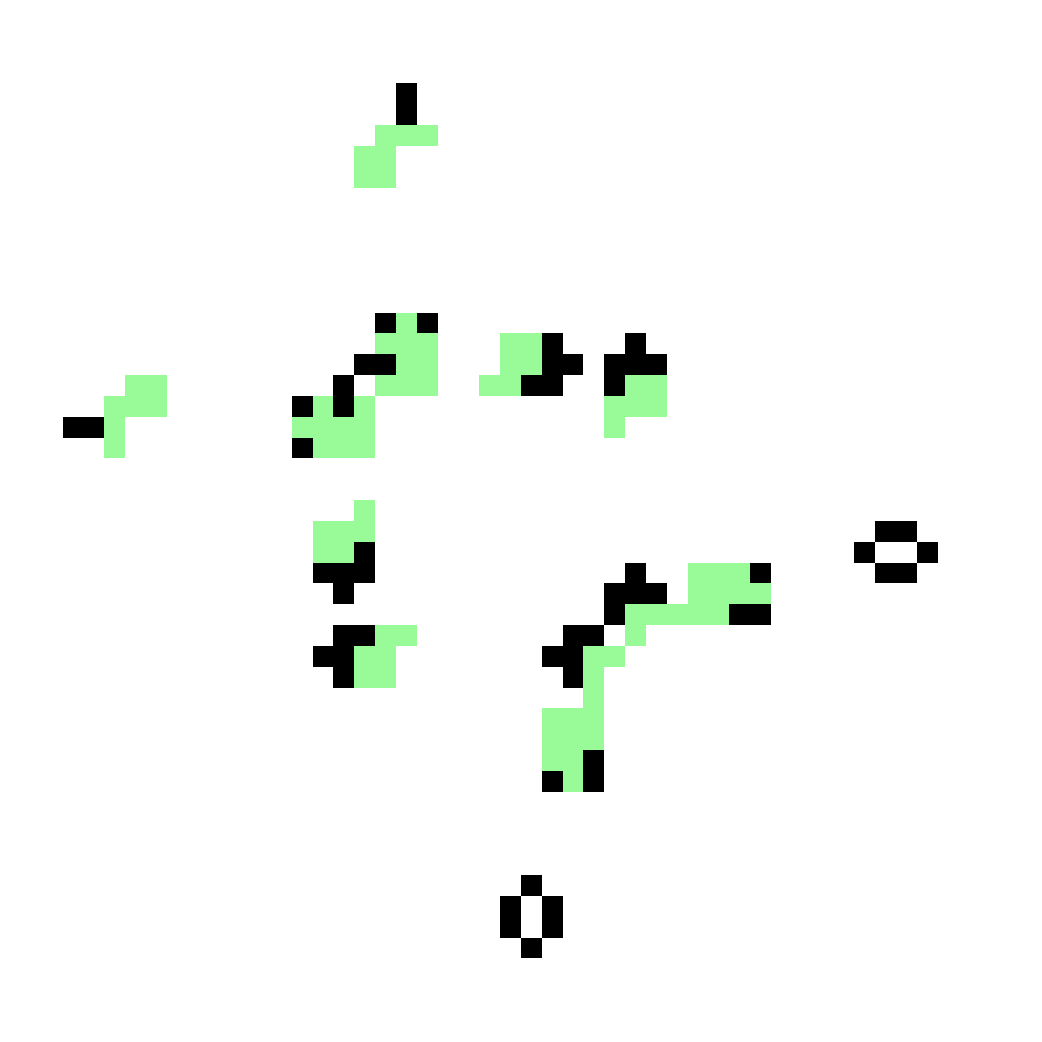}}
  \setlength{\fboxsep}{0pt}\fbox{\includegraphics[height=.25\linewidth]{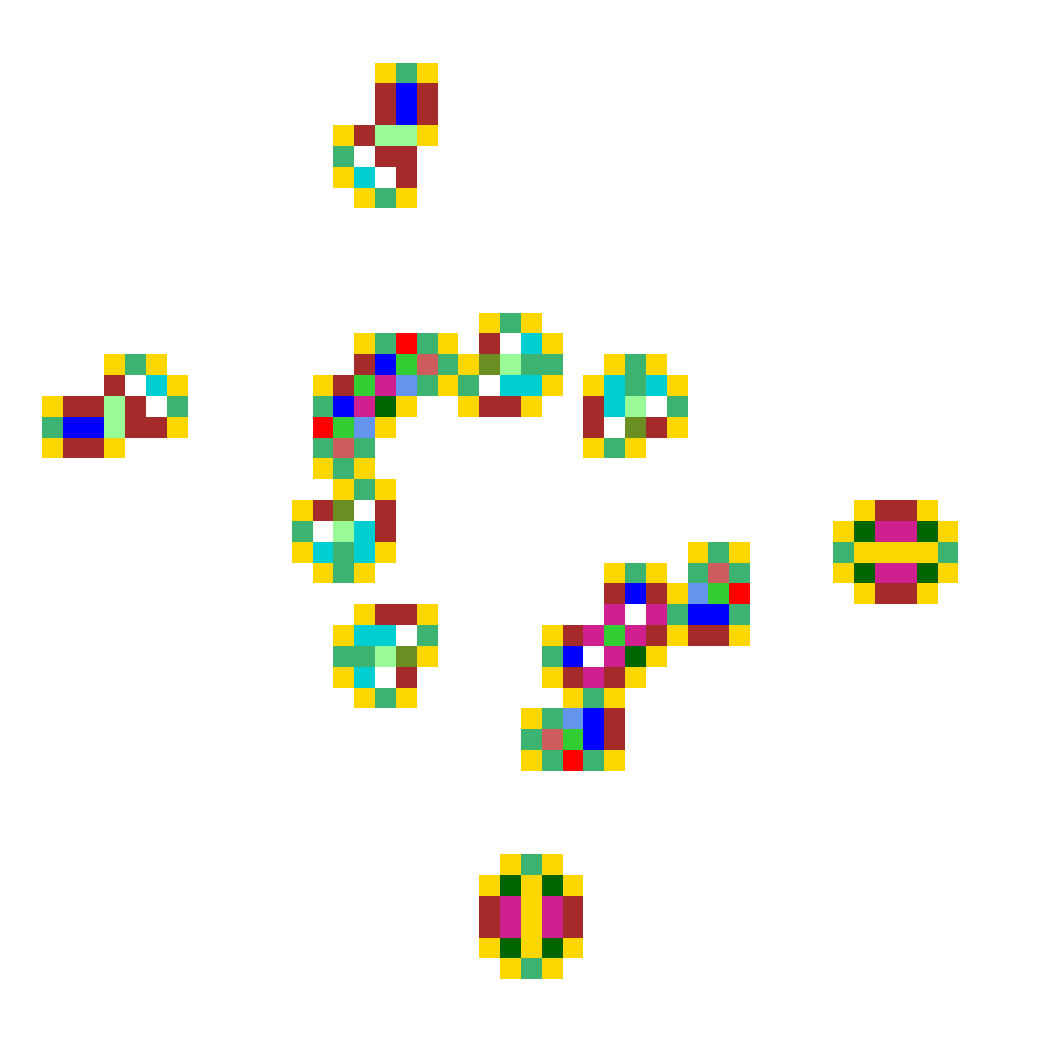}}
\phantom{i}
\raisebox{18ex}{
  \begin{minipage}[t]{.45\linewidth}
    {\footnotesize An intensive Ameyalli periodic space-time pattern \\
{\it far left}: colors by value with green \mbox{dynamic trails.}\\
{\it near left}: colors corresponding to IFH colors.}
  \end{minipage}
}\\
\begin{minipage}[c]{1\linewidth}
  \includegraphics[width=1\linewidth,height=.25\linewidth]{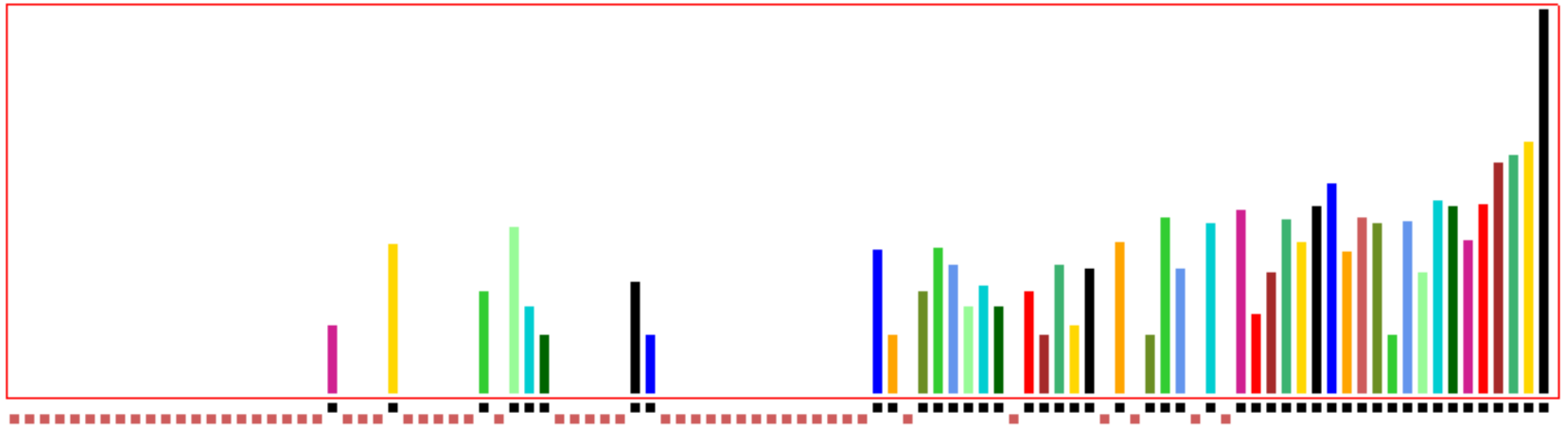}\\
  \includegraphics[width=1\linewidth,bb=6 13 835 25, clip=]{am_pdf-figs/am-iso}\\[-.5ex] 
  \includegraphics[width=1\linewidth,bb=6 13 835 25, clip=]{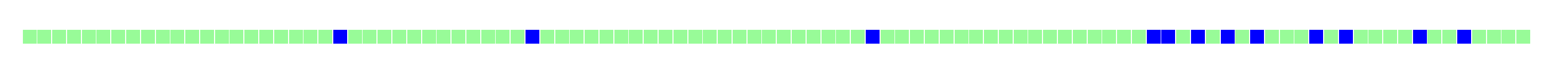}
\end{minipage}
}
\vspace{-1ex}
\caption[Ameyalli-rule idealized glider-gun IFH]{
  \textsf{The log$_2$ iso-rule IFH of the Ameyalli-rule driven by an ``intensive''
    periodic space-time pattern which includes all structures making logical gates.
    All 47 active bars were firstly filtered (black blocks), then all 53 neutral inputs were mutated
    to zero (red blocks), turning the Ameyalli-rule into its idealized form.
    The IFH and the intensive space-time pattern, and logical gates, are preserved, though other dynamics
    would be drastically altered.
    Below the IFH the corresponding iso-rule tables are shown, firstly the original Ameyalli with 31 inputs of 1,
    and below that the idealized Ameyalli with only 12 inputs of 1.\\
The iso-rules in hexadecimal hex are compared below:\\
{\small 26 42 a3 a9 08 8a 44 90 00 0b 54 50 90} ---Ameyalli iso-rule\\
{\small 00 00 01 00 08 00 00 10 00 03 54 50 90} ---idealized Ameyalli iso-rule\\
The idealized Hensel string translated for Golly\cite{Golly} is B2ci3ar5q/S01e2ek3qj4t5y
}}
\label{idealized iso-rule IFH}
\end{figure}

Here we construct the IFH compatible with the logical gates in section~\ref{Logical Gates}
in order to derive a stripped down ``idealized'' form of the logically universal Ameyalli iso-rule.
Employing the logical gates themselves is impractical because their dynamics is non-periodic.
Instead, in figure~\ref{idealized iso-rule IFH}  we construct an ``intensive'' periodic space-time
pattern that includes all the dynamical structures
that make the logical gates, with two interacting glider-guns to make mutually destructive
90$^0$ glider collisions, as well as gliders destroyed
by collisions with both eaters A and B.
This results in 47 active bars, but the majority are absent representing neutral inputs.
Any or all mutations to these neutral inputs has no affect on the ``intensive'' space-time
pattern, and when tested, the logical gates in section~\ref{Logical Gates} are preserved.
We suggest that if all these neutral inputs are set to zero, the resulting idealized iso-rule
will form an appropriate starting point for further Ameyalli studies.

\section{Summary and Discussion}
\label{Summary and Discussion}

The Ameyalli iso-rule is another example of the search for glider-guns,
then building logical gates.  Its spontaneously emergent glider-gun
was found from the input-entropy scatter-plot samples that favoured
both order and emergent gliders. Minor mutations created two types of
eater to stop the glider stream. With these ingredients, and by
adjusting collision dynamics according to precise timing and points of
impact, we were able to build the logical gates NOT, AND and OR
required for logical universality.

Further mutations would possibly uncover other artefacts of interest,
but for a better appreciation of the causal links between these
significant dynamical patterns and the responsible iso-groups inputs,
we applied the input-frequency histogram (IFH) method, which also
revealed neutral inputs where mutations have no effect.  As we have
shown for the Ameyalli, setting neutral inputs to zero relative to an
intensive glider-gun/eater system will reduce logically universal
iso-rules to their stripped down or ``idealized'' form where gliders,
eaters, glider-guns, and logical gates continue to be supported.  The
essential identity of Ameyalli, and other logically universal CA, are
these dynamical objects together with the significant part of the
iso-rule table that drives them. However, the possibility is there to
configure neutral inputs to create other unpredicted but relevant
structures.

An idealized logically universal iso-rule is the primitive of its huge
family of mutants that perform the same basic functions.  The
dynamics of idealized versions of all the logically universal
iso-rules mentioned in this paper, including the game-of-Life, are
worth investigating in further work.

\section{Acknowledgements}

The experiments were realised using DDLab, Mathematica and Golly.
Figures were made with DDLab.  This collaborative work began in 2017 at
a workshop in Ariege, France, and also at the Autonomous University
of Zacatecas, Mexico and in London, UK. J. M. G\'omez Soto
acknowledges his residency at the DDLab Complex Systems Institute and
financial support from the Research Council of M\'exico (CONACyT).

\end{document}